\documentclass[aps,pre,superscriptaddress]{revtex4-1}
\usepackage{graphicx}
\usepackage{color}
\usepackage{amsmath}
\usepackage{xfrac}
\usepackage{nicefrac}
\usepackage{mathrsfs,amsmath} 
\usepackage{amssymb}\usepackage{graphicx}
\usepackage{enumerate}
\usepackage{hyperref}
\usepackage{etoolbox} 
\usepackage{lipsum} 

\makeatletter
\appto{\appendix}{%
  \@ifstar{\def\theequation@prefix{A.}}%
          {}%
}
\makeatother

\def\_#1{^{}_{#1}}

\def\beq{\begin{equation}}\def\eeq{\end{equation}}
\def\bea{\begin{eqnarray}}\def\eea{\end{eqnarray}}

\begin{document}

\title{Modeling Spurious Forces on the LISA Spacecraft Across a Full Solar Cycle}
\author{Barrett M. Frank} \email{barrett.frank@montana.edu}\affiliation{Department of Physics,Grand Valley State University, Allendale, MI 49504  U.S.A.}\affiliation{Department of Physics, Montana State University, Bozeman, MT 59717-3840 U.S.A.}
\author{Brandon Piotrzkowski}\email{piotrzk3@uwm.edu}\affiliation{Department of Physics,Grand Valley State University, Allendale, MI 49504 U.S.A.}\affiliation{Department of Physics,University of Wisconsin - Milwaukee, Milwaukee, WI 53211 U.S.A.}
\author{Brett Bolen}\email{bolenbr@gvsu.edu}\affiliation{Department of Physics,Grand Valley State University, Allendale, MI 49504 U.S.A.}
\author{Marco Cavagli\`a}\email{cavagliam@mst.edu}\affiliation{Physics Department, Missouri University of Science and Technology, Rolla, MO 65409, U.S.A.}
\author{Shane L. Larson}\email{s.larson@northwestern.edu}\affiliation{Center for Interdisciplinary Exploration and Research in Astrophysics, Northwestern University, Evanston, IL 60208, USA}

\begin{abstract} \noindent
One source of noise for the Laser Interferometer Space Antenna (LISA) will be time-varying changes of the space environment in the form of solar wind particles and photon pressure from fluctuating solar irradiance. The approximate magnitude of these effects can be estimated from the average properties of the solar wind and the solar irradiance. We use data taken by the ACE (Advanced Compton Explorer) satellite and the VIRGO (Variability of solar IRradiance and Gravity Oscillations) instrument on the SOHO satellite over an entire solar cycle to calculate the forces due to solar wind and photon pressure irradiance on the LISA spacecraft. We produce a realistic model of the effects of these environmental noise sources and their variation over the expected course of the LISA mission.  

\end{abstract}
\maketitle
\section{Introduction}\label{sec.intro}
A major source of noise for any proposed space-borne gravitational-wave detector, such as the Laser Interferometer Space Antenna (LISA) \cite{LISAproposal}, or the proposed TianQin \cite{TianQin}, is the influence of space weather in the form of solar wind particles and photon pressure from solar irradiance\footnote{The term ``space weather'' is often focused on the local radiation environment of the Earth, but is used through out this paper in its more general sense, describing the time-varying interplanetary environment driven by changes in the solar wind and solar output.}. Since the Sun is a dynamical object, the space environment at any point in the solar system is expected to show fluctuations in time and magnitude on short timescales, as well as over the course of a full eleven-year solar cycle. The nominal LISA mission is 4 years for primary science operations with a reasonable expectation for a full mission lifetime of up to 10 years. Understanding the noise budget due to variations in solar wind output and irradiance over these timescales is important for science operations, data analysis, and the design of the spacecraft. 

Early tests of operations for the science measurement required by a single sciencecraft were conducted in situ, in the space environment, by the LISA Pathfinder mission, which made direct measurements of the charging of systems due to influx of high energy cosmic rays and solar wind particles \cite{PathfinderCharge}, and overall acceleration isolation of the gravitatioanl reference sensor \cite{PathfinderAccel}.

\begin{figure}[!htb]
     \centering
     \includegraphics[width=.5\linewidth]{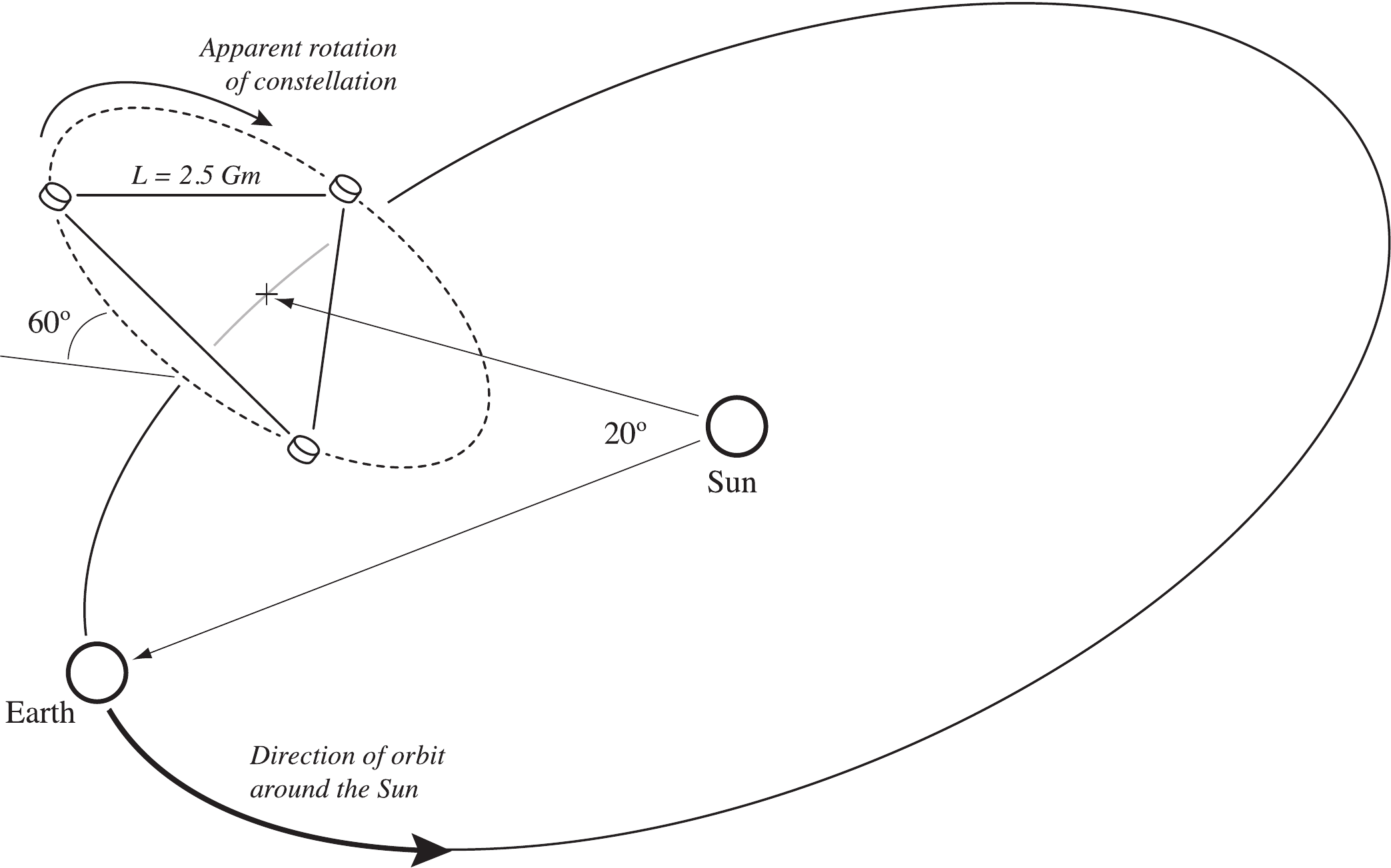}
     \caption{The location of the LISA constellation in relation to the Earth and the Sun \cite{LISAproposal}.  LISA forms an equilateral triangle with arms around 2.5 million km in length, inclined $60^{\circ}$ to the ecliptic, and trailing the Earth by $20^{\circ}$ in its orbit. To an observer located at the center of the constellation on the Earth's orbit, the configuration appears to rotate clockwise as viewed from the Sun.}\label{LISAoverview}
  
    
\end{figure}

LISA's nominal constellation configuration is shown in Fig.\ \ref{LISAoverview}. The center of the constellation will lie on the Earth's orbit, $20^{\circ}$ behind the Earth. Each LISA spacecraft will be on its own, individual Keplerian orbit around the Sun, but phased relative to the other two spacecraft. The constellation will appear to maintain its triangular shape to an inertial observer on Earth's orbit at the ``guiding center'' of the constellation, rotating clockwise around the guiding center as viewed from the Sun. Given this constellation configuration, the solar distance of each spacecraft will oscillate around 1 AU. Therefore, the LISA spacecraft are expected to experience space weather roughly comparable to Earth. 

The basic design of each LISA spacecraft consists of a free-falling test mass enclosed in an external structure shielding the test mass from external influences and containing the telescopes, lasers, optical benches, and other instrumentation. The spacecraft will include a flat solar panel facing the sun at all times. This panel will provide power to the spacecraft as well as protect the instruments from the solar wind particles and incident thermal radiation from the Sun. The spacecraft will respond to disturbances from the space environment shielding the test mass and allowing it to follow the (nominally) unperturbed motion of its space-time geodesic. Each LISA spacecraft will be equipped with a network of ``micro-Newton'' thrusters that will push the spacecraft in response to external forces. 

The magnitude of the forces due to solar wind and irradiance pressure can be estimated from the average properties of the Sun and solar wind parameters. The fluctuation of these forces on short timescales (that cause fluctuations in the measurement band) and their evolution over longer timescales (which affect mission performance on operations during different periods of observation) will be relevant to LISA measurements. The main goal of this paper is to use recent observations of the space environment (solar wind particle count and solar irradiance) taken over an entire eleven-year solar cycle to formulate a realistic picture of the space environment that a LISA spacecraft will experience over any expected mission lifetime.

In our analysis we use solar wind data from the Advanced Composition Explorer (ACE) satellite and solar irradiance data from the VIRGO (Variability of solar IRradiance and Gravity Oscillations) experiment aboard the ESA/NASA SOHO (Solar and Heliospheric Observatory) \cite{VIRGO1,VIRGO2,VIRGO3,ACE1}. ACE data provides solar wind velocity components and an alpha particle/proton ratio in 64 second intervals \cite{ACEdata}. For the solar irradiance analysis we use the 60-second averaged 1996-2014 data \cite{VIRGOdata}. Data from both experiments span more than one solar cycle and match LISA's $10^{-5}$ Hz to $0.1$ Hz nominal observational frequency. 

The ACE and SOHO satellites orbit around the Sun at the Earth-Sun L1 Lagrange point, $1.5 \times 10^{6}$  kilometers closer to the Sun than Earth. At the minimum radius of LISA's orbit, $r = 149.5 \times 10^{6}$ km and at the maximum LISA orbital radius, $r = 150.5 \times 10^{6}$ km, the discrepancy between the ACE and VIRGO readings (assuming that the particle flux follows an inverse square law) should be about 0.7\%. Thus these variations can be safely neglected when evaluating the effects of solar wind and irradiance pressure over LISA's orbit, mostly since this variation occurs at a frequency lower than detectable by LISA.

This paper is organized as follows. In Sec.\ \ref{sec.SolarWind} we calculate the force on the LISA spacecraft due to the solar wind. In Sec.\ \ref{sec.Irradiance} we calculate the force due to solar irradiance. Conclusions are summarized in Sec.\ \ref{sec.conclusion}. 

\section{Force due to Solar Wind}\label{sec.SolarWind}

The ACE dataset provides particle density and velocity of alpha particles/protons in the solar wind. This dataset can be used to calculate the force exerted on a LISA spacecraft, starting with the number of particles $N$ per unit time colliding with the satellite 
\begin{equation}
N = n \, v\, A \,\cos{\phi},
\end{equation}
where $n$ is the particle number density (measured by ACE), $v$ is the wind speed, $A$ is the area of the LISA solar array, and $\phi$ the angle between the normal of the array and the orbital plane.  
The solar wind has a velocity that is not necessarily directed on a line from the Sun to the spacecraft, resulting in a net force that may point in a different direction.

\begin{figure}[!htb]
\includegraphics[width=0.4\textwidth]{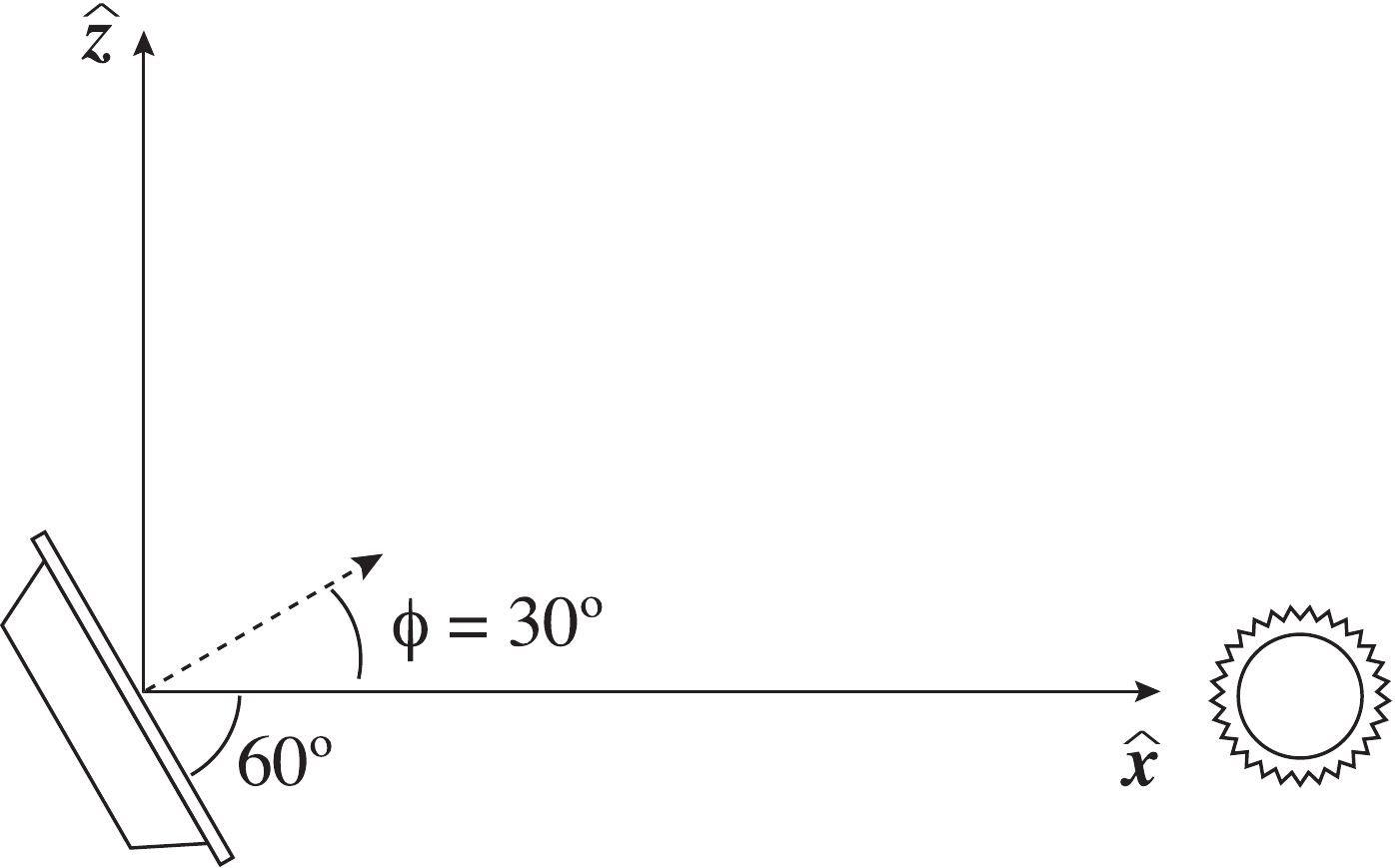}

\caption{Geocentric solar ecliptic (GSE) coordinates.}\label{GSE}
\end{figure}

To define the force on the spacecraft we use a geocentric solar ecliptic (GSE) coordinate system with the $x$ axis pointing from the satellite toward the Sun and the $z$ axis pointing along the normal of the orbital plane (see Fig.\ \ref{GSE}). 

The force due to the particle collisions is
\bea
F_{x} &=& (N_{p}m_{p} + N_{\alpha}m_{\alpha})[(1+R\cos(2\phi))v_{x}+R\sin(2\phi)v_{z}]\,, \nonumber \\
F_{y} &=&(N_{p}m_{p} + N_{\alpha}m_{\alpha})[(1-R)v_{y}]\,, \nonumber \\
F_{z} &=& (N_{p}m_{p} + N_{\alpha}m_{\alpha})[(1+R\cos(2\phi))v_{z}+R\sin(2\phi)v_{x}] \,,
\label{forcewind}
\eea
where ${\vec v} = \{v_x,v_y,v_z\}$ is the particle velocity, $N_{p}$ and $N_{\alpha}$ are the rates of proton and alpha particle collisions with the satellite, $m_{p}$ and $m_{\alpha}$ are the proton mass and alpha particle mass, respectively, and $R$ is the fraction of particles which are reflected off the satellite.  In the following, we assume $R = 1$, i.e., that all particles are perfectly reflected off the LISA Solar Array as a worst-case scenario. Under this assumption, Eqs.\ (\ref{forcewind}) simplify to
\bea
F_{x} &=& (N_{p}m_{p} + N_{\alpha}m_{\alpha})[(1+\cos(2\phi))v_{x}+\sin(2\phi)v_{z}]\,, \nonumber \\
F_{y} &= &0\,, \nonumber \\
F_{z} &=& (N_{p}m_{p} + N_{\alpha}m_{\alpha})[(1+\cos(2\phi))v_{z}+\sin(2\phi)v_{x}]. 
\eea


The ACE duty cycle is 67.5\% over the analyzed 11-year period, with data gaps appearing at irregular intervals. Most of these gaps are relatively short in duration, consisting of only 1 or 2 consecutive missing values and showing very small fluctuations in the values over the length of the gap. Some of these gaps are from known spacecraft processes. For instance, a time series analysis of the distance between gaps shows that there are at most 30 consecutive data readings and that every $31^{\rm st}$ reading is a gap. These missing data are due to ACE recalibration of the nominal mode operation algorithm \cite{ACE1}.  This process requires full range readings which have a reduced energy resolution and thus are not included in the public ACE data. To address the effect of data gaps in our analysis to eventually calculate a Fourier transform, we have developed different approaches to mitigate the effect of their presence on our results.



We classify the gaps in three categories according to their length and employ separate techniques to reconstruct the missing data: Single-value gaps (Type A), gaps ranging in length between 2 and 24 data readings (Type B), and gaps of length 25 or more (Type C). Type A data are reconstructed by a simple linear interpolation, i.e., by replacing the missing data with the average of the data values on either side of the gap. This method works quickly and efficiently and does not introduce any spurious spectral artifact on the signal. Type B gaps are filled using a linear interpolation model with added Gaussian noise. To insure that the Gaussian noise has the same statistical character of the surrounding data, $\sigma$ is determined by sampling continuous stretches of data on either side of the gaps. Each pre- or post-gap stretch of data are taken to be equal in length to the gap and used to calculate individual standard deviations $\sigma_{1}$ and $\sigma_{2}$.  For the $i$th entry in a Type B gap of length $l$, where $1\leq i \leq l$, $\sigma_{1}$ and $\sigma_{2}$ are then weighted based on how close the entry is to each stretch as
\begin{equation}
\sigma = \frac{l-(i-1)}{l+1}\sigma_{1}+\frac{i}{l+1}\sigma_{2}\,.
\end{equation}

Attempts at simulating data in longer gaps will necessarily introduce spurious noise that will show up in a spectral analysis. Therefore, type C gaps are managed by windowing the data around the gaps with a function which slowly falls off to zero, reducing high frequency Gibbs phenomena in the spectrum caused by sharp discontinuities around gaps. The 25 data values on either side of Type C gaps are multiplied by half of a Hann window
\begin{equation}
w(n) = 0.5\left[1-\cos{\left(\frac{2\pi n}{N-1}\right)}\right]\,,
\end{equation}
where $N=51$ and $0<n<N-1$.

\begin{table}[h] 
\centering
\begin{tabular}{|c|c|c|c|}
\hline
Type & Gap Length & Total percentage of data & Filling Method \\
\hline
A & 1 &$\approx 87.8\%$ &Linear Interpolation\\
B & 2 - 24 & $\approx 11.3$ \% & Linear Interpolation + Gaussian Noise \\
C & 25 + & $\approx 0.9$& Window Edges of Gap \\
\hline
\end{tabular}
\caption{Characterization of gaps based on length and filling technique for the solar wind using the 1999 ACE data set. The statistics are similar across any long span of the ACE data.}\label{gaptable}
\end{table}

\begin{figure}[h]
\includegraphics[width=0.45\textwidth]{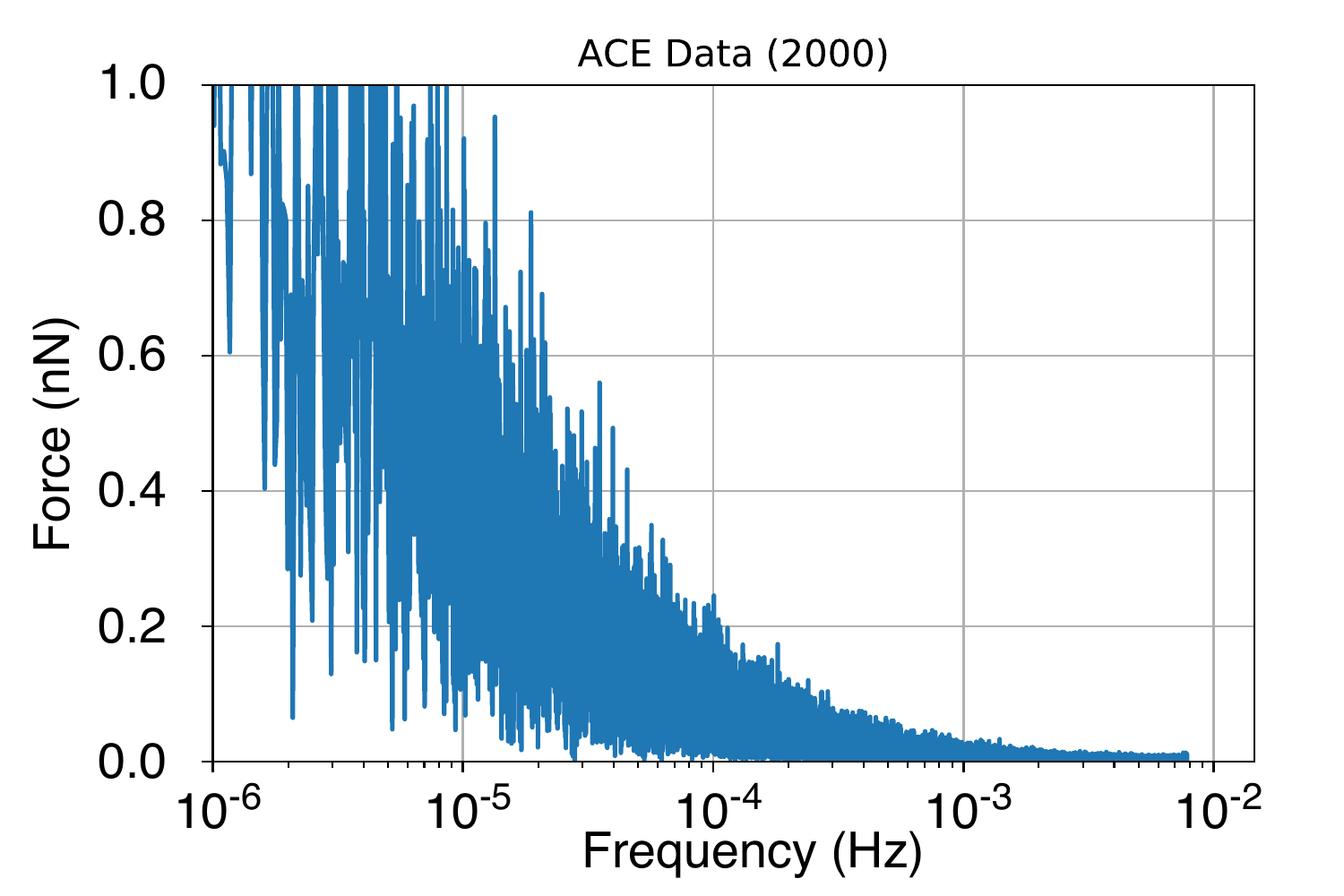}
\includegraphics[width = 0.45\textwidth]{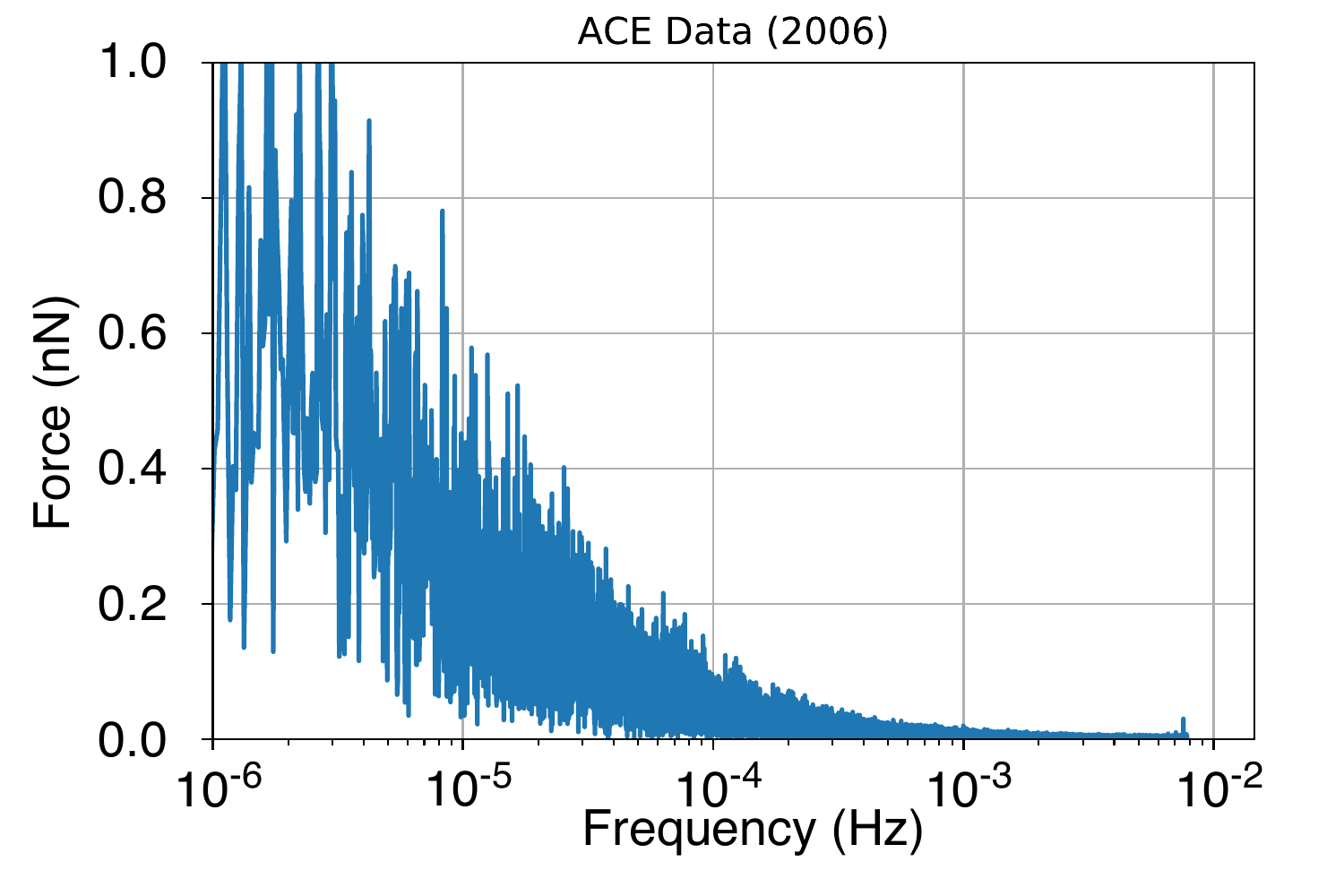}
\caption{Plot of force due to solar wind as measured by the ACE spacecraft vs. frequency in LISA sensitivity band for the years 2000 (left) and 2006 (right). Note the difference is due to 2000 being near solar maximum and 2006 is a solar minimum.  Note that the years of these plot are the same as was done for solar irradience (see figure \ref{VirgoFFT})
}\label{fig:acefft}
\end{figure}

Utilizing this approach to handling data gaps, we have analyzed the full span of ACE data (more than a complete solar cycle). Our results are shown in Fig.\ \ref{fig:acefft} for a solar maximum period (year 2000), when the Sun is very active and the solar wind is more dynamic, and for a solar minimum period (year 2006) when the solar activity and solar wind are much more quiescent. If future solar cycles show similar activity, we expect this to be a good representation of the influence of the solar wind on LISA during its mission lifetime. The results show fluctuating forces at different frequencies with stronger peaks at lower frequencies, and a greater (smaller) amplitude of the force during the solar maximum (minimum). During the complete solar cycle covered by ACE, had LISA been flying, the force due to the solar wind would never have exceeded 250 nN, a value well within the requirements for LISA. Assuming that future solar cycles will not differ much from ACE observations, the solar wind should not be a significant noise source for the LISA mission.

It should be noted that solar energetic particle events may also influence LISA in other ways, most notably spacecraft charging. This is expected to produce noise on LISA as charges bound to the spacecraft or in the test masses moving through the interplanetary magnetic field would experience a Lorentz force.

\section{Force by Solar Irradiance}\label{sec.Irradiance}

To calculate the force of solar radiation on the LISA spacecraft we consider the standard LISA design with a flat solar array which is always facing the Sun at a constant angle. The results presented below are specific to the exact current LISA configuration, they are easily adapted to LISA's ultimate spacecraft design and profile. Two factors contribute to radiation pressure on the panel: the direct force from the incident radiation (absorption and reflection) and the resulting back reaction due to thermal reradiation. 

We assume the solar array to be made of two different materials with different optical properties: solar cells and optical solar reflectors. The solar cells may be treated as blackbody objects since they absorb light efficiently across the solar frequency spectrum. The optical solar reflectors are designed to reflect light to reduce the overall temperature of the panel and are placed where solar cells do not fit. Typical solar arrays are  manufactured with a cover glass that rejects frequencies outside the effective range of the solar cells (wavelengths from smaller than \(\sim 300\) nm and larger than \(\sim 1350\) nm) \cite{coverGlass}.  This effect may be computed by using Planck's radiation law. The solar array is designed to be thermally isolated from the rest of the components. If this was not the case, the solar array would create temperature gradients within the spacecraft producing additional noise on the test mass. The thermal noise allowed in the spacecraft noise budget on the test masses is low (\cite{LPPA,YellowBook2}), so we only consider the top side of the panel when calculating the force by re-radiation.

\subsection{Incident Solar Radiation}\label{sub.SolarRadiation}
Since photons are largely unaffected by magnetic fields, we expect in the purely absorptive case the force of incident radiation in GSE coordinates to be 
\begin{equation}
\vec{F_I} = -\frac{E_f A }{c} \hat{x}\,, \label{zeroirradforce}
\eeq
\noindent where \(E_f\) is the solar irradiance (W/m$^2$),  and $A$ is the effective cross-sectional area of the solar array \cite{Force}, and ${\hat x}$ is the unit vector in the orbital plane defined in Figure \ref{GSE}.

In our analysis we consider a modern coverglass with lower and upper wavelength \(\nu\) cutoff of $300$ nm and $1350$ nm, respectively. Using Planck's law, the fraction of absorbed power by the panel is

\beq
\delta =  \frac{ \displaystyle \int_{\eta_{uv}}^{\eta_{ir}}\frac{\eta^3 d \eta}{e^{\eta}-1} }{\displaystyle  \int_{0}^{\infty}\frac{\eta^3 d \eta}{e^{\eta}-1}}  =  \frac{ \displaystyle \int_{\eta_{uv}}^{\eta_{ir}} \frac{\eta^3 d \eta}{e^{\eta}-1} }{\displaystyle \frac{\pi^4}{15} }\approx  .81 \,,\notag 
\eeq
\noindent  where $\eta= \hbar \nu/k_b T$, \(\hbar\) is the reduced Planck's constant, \(k_b \) is the Boltzman constant, and $T$ is the surface temperature of the Sun.  As stated above the normal vector to the solar array is inclined by an angle of $\phi = 30^{\circ}$ from the plane of the solar system. Now if we include reflection and the properties of two materials of the solar array, the total force is
\beq
\vec{F_I}=\begin{bmatrix} F_{I,x} \\ F_{I,z}\end{bmatrix}= -\frac{E_f}{c} \left( A_{sc} \begin{bmatrix} 1+ (1-\delta \alpha_{sc}) \cos{2 \phi} \\ (1- \delta \alpha_{sc})  \sin{2 \phi} \end{bmatrix} + A_{osr} \begin{bmatrix} 1+ (1-\delta \alpha_{osr}) \cos{2 \phi} \\ (1- \delta \alpha_{osr})  \sin{2 \phi} \end{bmatrix} \right) \label{reflectforce}~.
\eeq
\noindent Here $\alpha_{sc} = 0.9$ and $\alpha_{osr}=0.08$ are the absorption coefficients of the solar cells and the optical solar reflectors, respectively, while $A_{sc}=3.0$ m$^2$ and $A_{osr}=0.3$ m$^2$ are the respective areas \cite{coverGlass}. Note  that  $2 \phi$  is  used since  by  rotating  $45^\circ$  all  reflected  irradiance  goes entirely in the z-direction.

\subsection{Force by Re-radiation}\label{sub.reradiation}
Photons being re-radiated by the solar array  can be modeled via blackbody radiation in an isotropic fashion with a power determined via the  Stefan-Boltzmann law modified by the emissivity $\epsilon$:  $E_F= \epsilon \sigma T^4 $. 
To calculate the force re-radiated from the solar array, we first find the steady-state temperature of the solar array. This temperature \(T_{sat}\) is found by equating the power entering the array to the power being re-radiated
\beq
0 = \left( \delta \alpha_{sc} A_{sc}  + \delta \alpha_{osr} A_{osr} \right) E_f - \left( A_{sc} \epsilon_{sc} \sigma +A_{osr} \epsilon_{osr} \sigma \right) T_{sat}^4\,, \notag
\eeq
\noindent where \(\epsilon_{sc} =0.86\) and  \(\epsilon_{osr}=0.86\) are the respective emissivities of the solar cells and the optical solar reflectors. Solving for the satellite temperature we have
\beq
T_{sat} = \left( \delta \frac{\alpha_{sc}A_{sc} +\alpha_{osr} A_{osr}}{\epsilon_{sc}A_{sc}+\epsilon_{osr}A_{osr}} \frac{E_f}{\sigma} \right)^{1/4}~. \notag
\eeq
Since black bodies re-radiate isotropically, the net force is reduced by a factor of \(1/ \pi\). Following the Stephan-Boltzmann Law, the contribution of re-radiation to the force on the spacecraft is
\begin{align}
\vec{F_R} = \begin{bmatrix} F_{R,x} \\ F_{R,z} \end{bmatrix} &= - \frac{1}{\pi}\left(\epsilon_{sc}A_{sc}+\epsilon_{osr} A_{osr}\right) \frac{\sigma T_{sat}^4}{c} \begin{bmatrix} \cos{\phi} \\ \sin{\phi} \end{bmatrix} \notag \\
 &= - \frac{1}{\pi}\frac{\delta E_f}{c} \left(\alpha_{sc}A_{sc}+\alpha_{osr} A_{osr}\right) \begin{bmatrix} \cos{\phi} \\ \sin{\phi} \end{bmatrix}\,.  \label{irradforce}
\end{align}

\subsection{Irradiance Data from VIRGO}\label{sub.virgo}




Just as with the ACE data, an operational difficulty with the VIRGO data for this study is the presence of many missing data points, or ``gaps'', in the data set. These gaps come from periods where the instruments were not taking data due to operational glitches, or the data was considered to be unreliable by the VIRGO research team.  Analysis of these gaps shows the vast majority of gaps are small, with long stretches of contiguous data (typical data spans are around 300-700 data points in length before encountering a gap). A graphical analysis of these gaps is shown in Figure \ref{fig:gap1}.
\begin{figure}[h]
\includegraphics[width=0.45\textwidth]{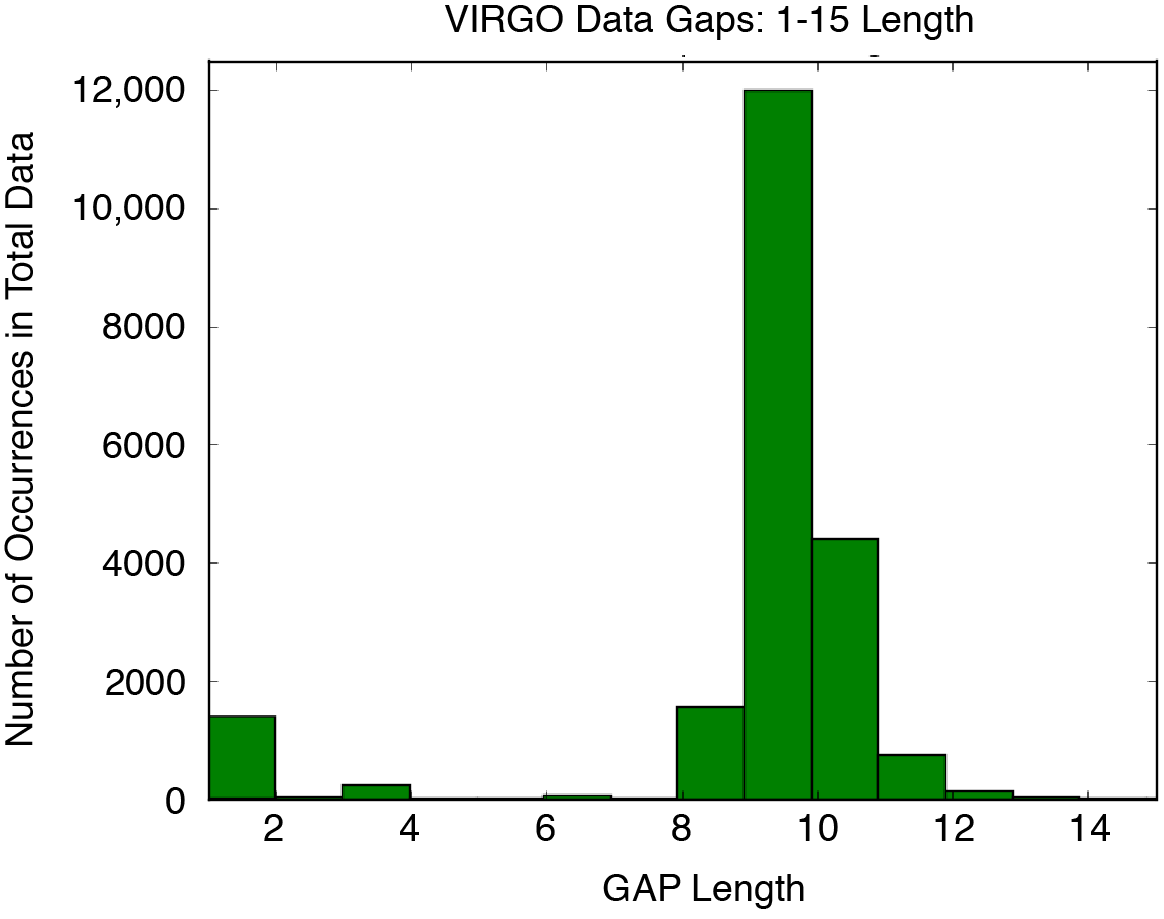}
\includegraphics[width=0.45\textwidth]{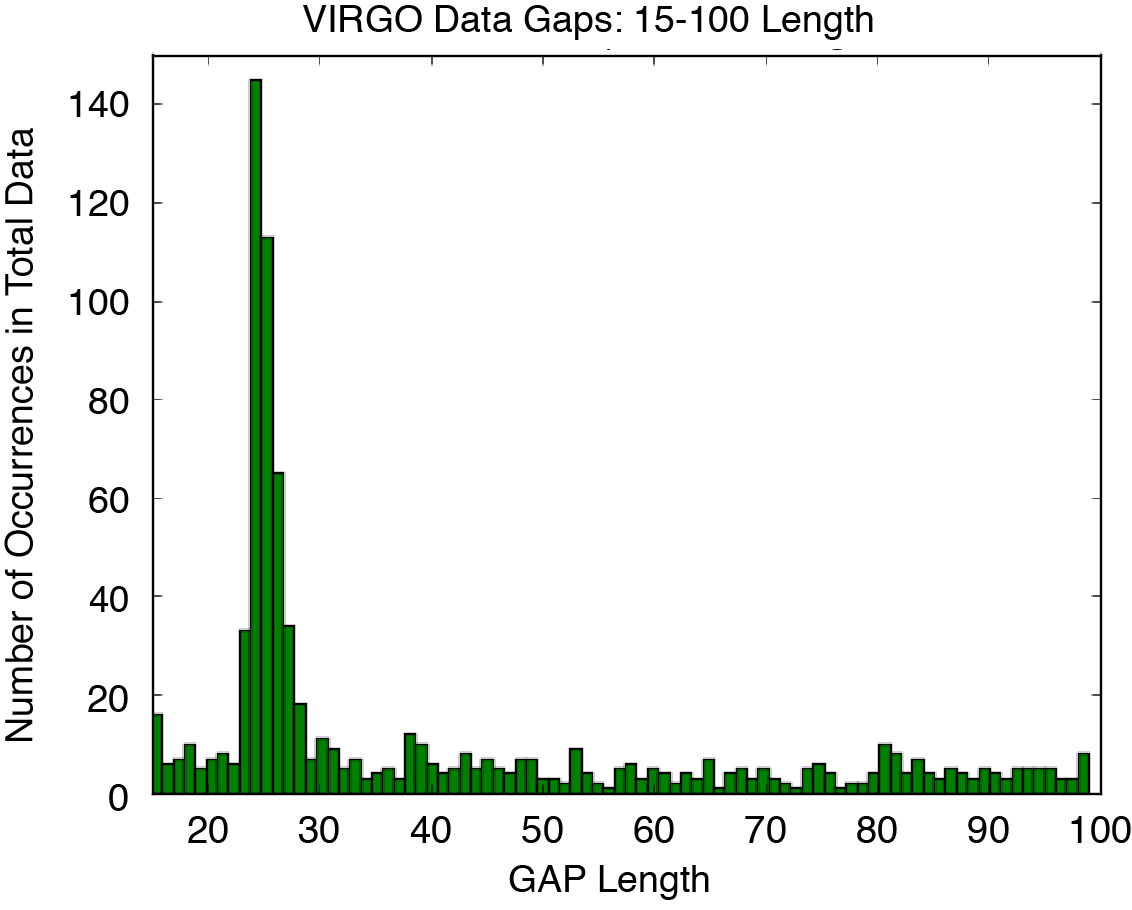}
\caption{Number of gap occurrences as function of the gap length in VIRGO data.\label{fig:gap1}}
\end{figure}

Ignoring the data gaps is expected to produce spurious features in the noise spectrum. Due to nature of VIRGO's gaps, the procedure of reconstructing the missing data is more complex than for ACE data. Thus we adopt a different gap-filling procedure based on machine learning. This technique allows us to reconstruct the missing data in a way which is consistent with the local characteristics of the irradiance data in the neighborhood of each gap. The algorithm first removes the DC offset of the data, and then reconstructs the data with a Gaussian Process (GP) algorithm. GPs are powerful generic supervised machine learning methods designed to solve regression problems. The advantages of GP algorithms is their ability of interpolating the observations in a probabilistic way while allowing for different relationships between data points (the ``kernel'') to be specified.

In our analysis, we use the GP Scikit implementation \cite{scikit} based on algorithm 2.1 of \cite{GPref} with a mixture of four kernels: (1) A stationary Radial-Basis Function kernel with default length scale parameter and length scale bounds ($10^{-5}$, $1$); (2) A Rational Quadratic kernel with default scale mixture parameter and bounds, length scale equal to 100 and bounds ($10$, $10^4$); (3) An ExpSineSquared kernel with gain equal to 10, default periodicity with bounds ($10^{-2}$, $10^5$), length scale equal to 1000 and default bounds; and finally (4) a White kernel with default noise level and bounds ($10^{-10}$, $10^{-6}$) which is used to estimate the noise level of the sample.

We test the reconstruction algorithm on an artificial data set as follows. We take the longest stretch of continuous data (1664 samples) in the full set of data from 1995 to 2014 and
insert an arbitrary number of fake gaps of different durations at random positions. The Amplitude Spectral Density (ASD) of the reconstructed data is then compared to the ASD of the
original data. Figure \ref{fig:fake} shows the difference of the two ASDs when four gaps with random duration from 1 to 10 samples are inserted in the data. The reconstructed ASD agrees
with the original ASD within a few percent level in the frequency range between $10^{-3}$ Hz and $8\cdot 10^{-3}$ Hz, showing that the algorithm does not introduce and significant
spurious lines. 

\begin{figure}[h]
\includegraphics[width=0.75\textwidth]{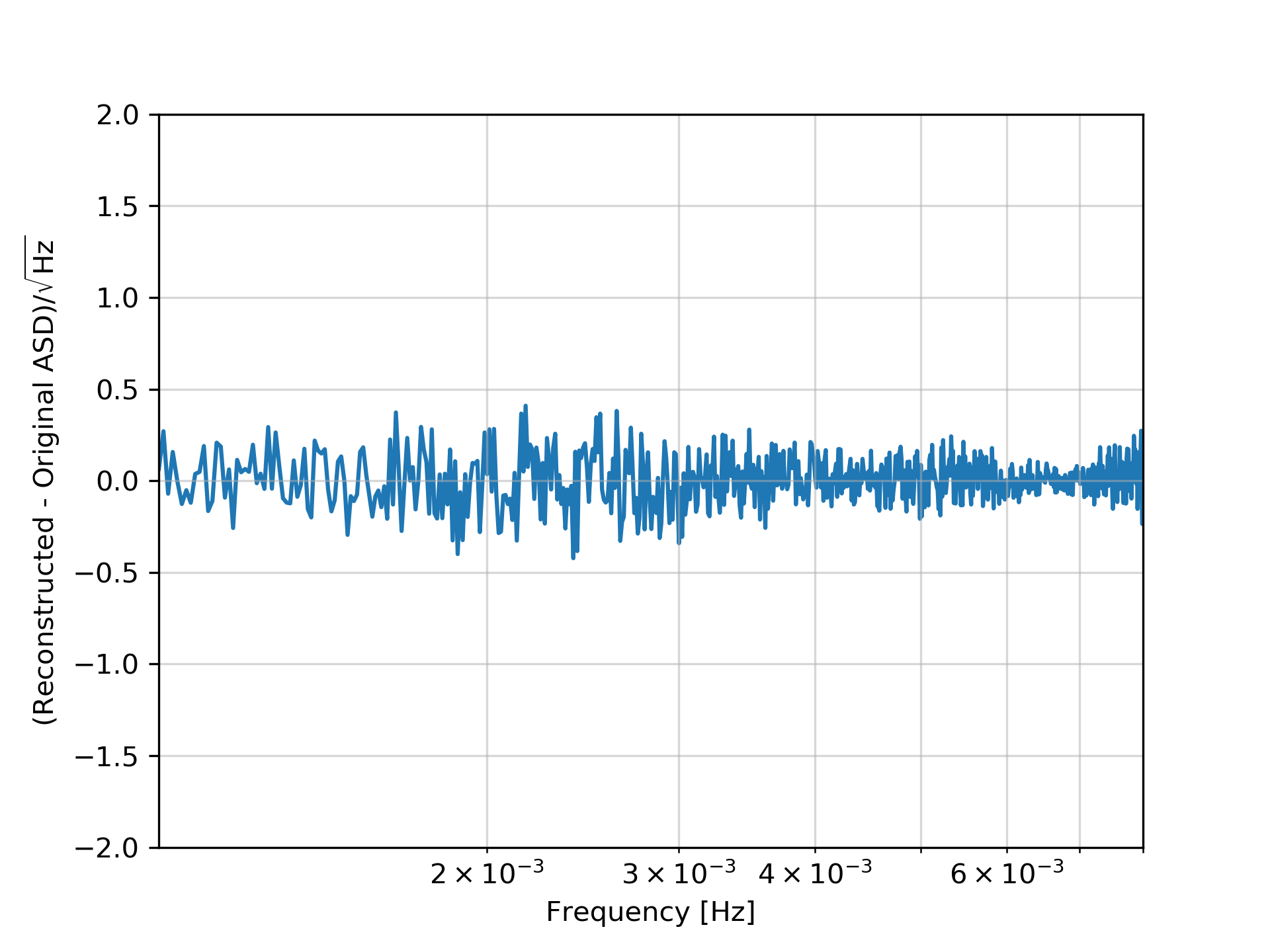}
\caption{Difference of the reconstructed ASD for a stretch of 1664 continuous data samples, where four artificial data gaps of different duration have been inserted at random positions, and the ASD from the original data. The comparison shows that the machine learning algorithm does not introduce any spurious lines. \label{fig:fake}}
\end{figure}

After data reconstruction, we reduce spurious red-noise in this procedure by filtering with a 4-th order Butterworth lowpass filter below $0.008$ Hz. We also apply a 2-nd order Butterworth bandpass filter with notch widths of $6\cdot 10^{-6}$ Hz and $4\cdot 10^{-4}$ Hz to remove artifacts due to the highest-frequency instrument noise in the 1-minute VIRGO data and its sub-harmonics at 0.005556 Hz and 0.002778 Hz, respectively.

\subsection{Results for Solar Irradiance Data}\label{sub.irradianceData}

\begin{figure}[h]
\includegraphics[width=0.75\textwidth]{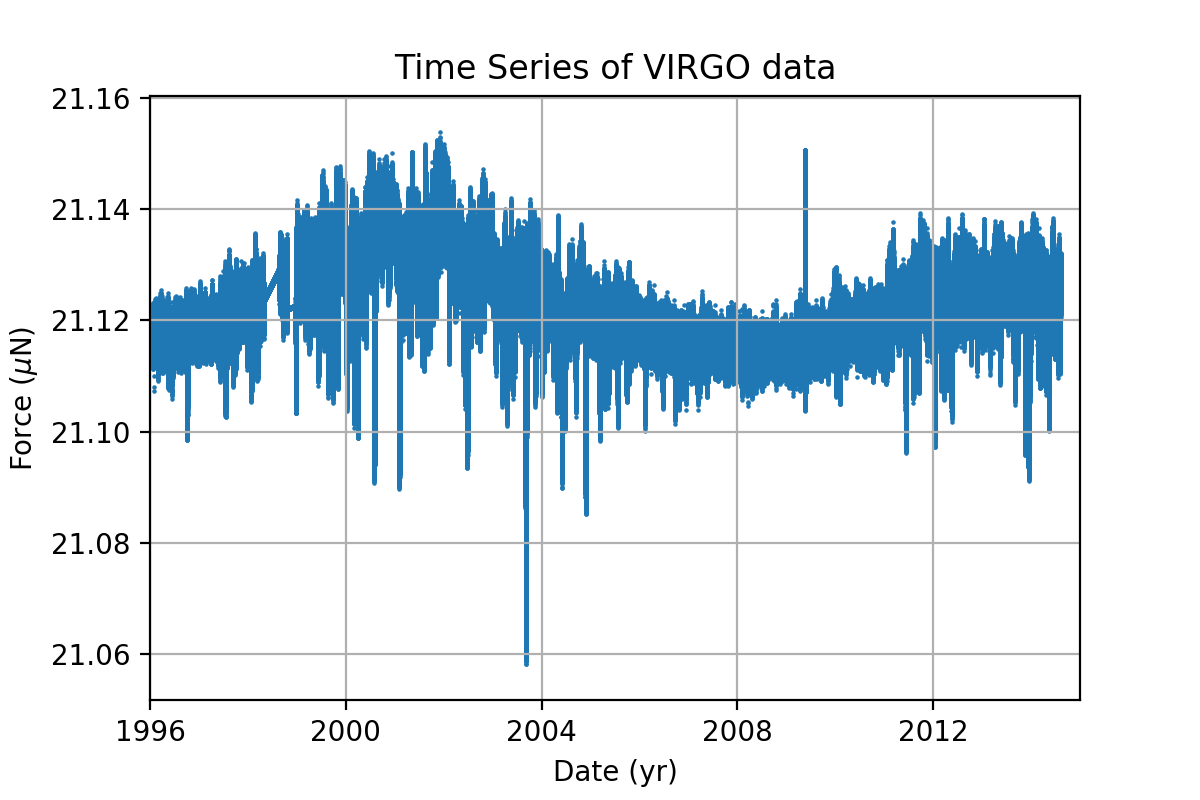}
\caption{Time series of the irradiance force experienced by the LISA spacecraft from the entire VIRGO dataset. The 11-year peak-to-peak variation in the force is a manifestation of the solar activity cycle. 
}\label{Virgoallyrs}
\end{figure}

Applying our force analysis to the entire VIRGO data set results in the force time series shown in \ref{Virgoallyrs}. There is a marked variation with solar cycle, with fluctuations around the mean force signal. The irradiance force stays within a narrow range and is well below the $100 \mu$N limit of the LISA thrusters \cite{FTR}. We should note the maximum difference of irradiance from the mean due to the solar variations is about \(.3\%\), which is less the \(.7\%\) variation predicted from the orbit.

\begin{figure}
\includegraphics[width=0.45\textwidth]{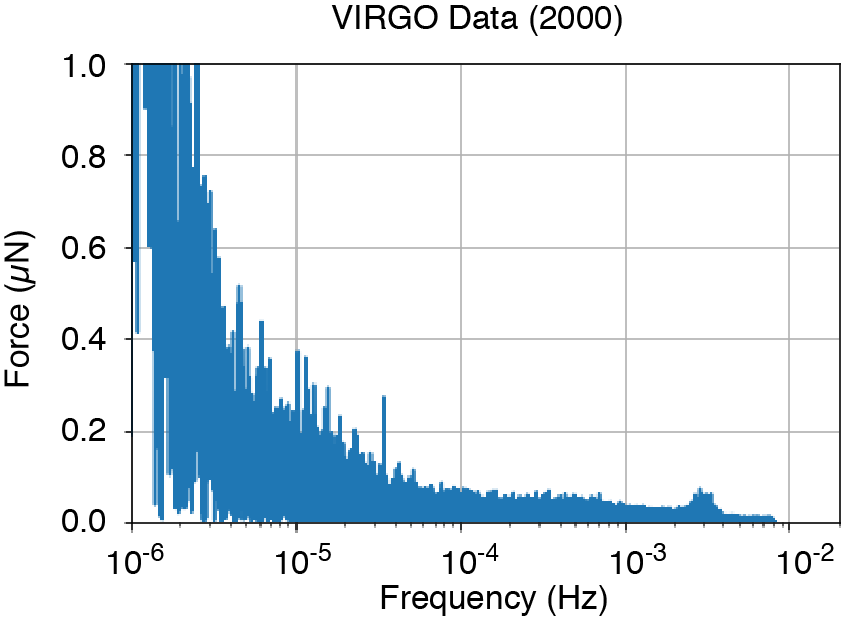}
\includegraphics[width=0.45\textwidth]{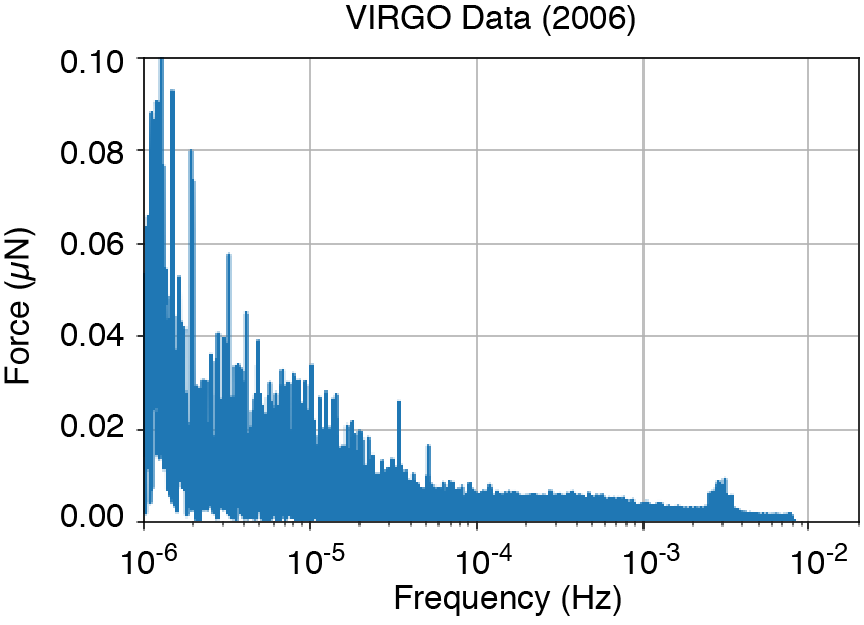}

\caption{Fourier transform of the force calculated from the VIRGO dataset within the LISA bandwidth for the years 2000 and 2006. The ``bump'' at 0.003 Hz is due to the 5-minute solar oscillation.
}\label{VirgoFFT}
\end{figure}

The Fourier transform of the irradiance force is shown in Figure \ref{VirgoFFT}. There is a strong low-frequency peak, associated with the solar cycle variation. At higher frequencies that overlap with the LISA band, there is a bump around $0.003$ Hz, as well as some prominent lines between $10^{-5}$ and $10^{-4}$ Hz, the signature of different pressure modes of the Sun creating fluctuations in the irradiance. Parts of the surface of the Sun move up and down, increasing and decreasing solar irradiance accordingly. Our results for this section are consistent with VIRGO dataset analysis in the solar literature \cite{DIS}. The ``bump'' around $0.003$ Hz which is the known solar ``5-minute oscillation''.   


\section{Conclusion and Future Work}\label{sec.conclusion}

In this paper we have considered the archived data from a complete solar cycle as recorded by ACE and VIRGO, and used these data to model the effects that would be experienced by a single LISA spacecraft. The goal was to understand the overall spectrum of external noise influences, and to better understand transient excursions that might occur due to bursts in solar activity or unusually strong solar events. Given the data recorded over the last solar cycle, a similar variance in solar activity during the LISA mission will be well within the planned capabilities of the planned LISA flight thruster systems, and will not produce untoward amounts of noise in the LISA measurement band. 

In principle, the environmental effects on LISA should be combined (irradiance and solar wind) to find the total contribution to the LISA noise, but given the disparity between the sample rates and differing gap structure in ACE and VIRGO, such a combination is outside the scope of this paper. However the independent analysis of each effect presented here is well justified since, as expected, the force due to solar irradiance dominates over the force due to the solar wind. The overall level of the forces do fluctuate with solar cycle, peaking during solar maximum, but then falling during solar minimum. Such considerations maybe be important when ultimately considering the flight dates of a spaceborne gravitational wave observatory, but should not impact the decision for performance at the required LISA levels. 

Studies like the one presented here are also of interest to the current generation of LISA modeling, where high fidelity end-to-end simulations of the LISA noise and LISA data are being produced as part of the LISA Consortium's effort to begin developing the complex pipelines that will be needed for LISA data analysis.

There are some questions that have not been answered as part of this study, but may be interesting in future work. Foremost would the aforementioned influence of solar activity on spacecraft charging rates. Spurious charge is expected to be a source noise that is mitigated by an onboard charge management system which expulses accumulated charge via photoelectric ejection; significant variations in the charging rate must be able to be addressed by the performance specs of this system. Since the solar wind is one source of charged particles (in addition to high-energy cosmic rays), it would be interesting to build a higher fidelity model of the spacecraft-solar wind interaction that accounts for embedding and penetration of high energy particles, rather than using the hard-sphere collision model employed here. It would also be of interest to consider whether or not LISA housekeeping data, in particular the spacecraft's thermal state and its thruster record, could be used as the input for an inverse problem to describe the space weather environment around LISA \cite{shaul2006solar}. This would provide multi-point measurements of the solar output which might plausibly be useful to the heliophysics community. 

\bibliography{swReferences}

\begin{thebibliography}{20}%
\makeatletter
\providecommand \@ifxundefined [1]{%
 \@ifx{#1\undefined}
}%
\providecommand \@ifnum [1]{%
 \ifnum #1\expandafter \@firstoftwo
 \else \expandafter \@secondoftwo
 \fi
}%
\providecommand \@ifx [1]{%
 \ifx #1\expandafter \@firstoftwo
 \else \expandafter \@secondoftwo
 \fi
}%
\providecommand \natexlab [1]{#1}%
\providecommand \enquote  [1]{``#1''}%
\providecommand \bibnamefont  [1]{#1}%
\providecommand \bibfnamefont [1]{#1}%
\providecommand \citenamefont [1]{#1}%
\providecommand \href@noop [0]{\@secondoftwo}%
\providecommand \href [0]{\begingroup \@sanitize@url \@href}%
\providecommand \@href[1]{\@@startlink{#1}\@@href}%
\providecommand \@@href[1]{\endgroup#1\@@endlink}%
\providecommand \@sanitize@url [0]{\catcode `\\12\catcode `\$12\catcode
  `\&12\catcode `\#12\catcode `\^12\catcode `\_12\catcode `\%12\relax}%
\providecommand \@@startlink[1]{}%
\providecommand \@@endlink[0]{}%
\providecommand \url  [0]{\begingroup\@sanitize@url \@url }%
\providecommand \@url [1]{\endgroup\@href {#1}{\urlprefix }}%
\providecommand \urlprefix  [0]{URL }%
\providecommand \Eprint [0]{\href }%
\providecommand \doibase [0]{http://dx.doi.org/}%
\providecommand \selectlanguage [0]{\@gobble}%
\providecommand \bibinfo  [0]{\@secondoftwo}%
\providecommand \bibfield  [0]{\@secondoftwo}%
\providecommand \translation [1]{[#1]}%
\providecommand \BibitemOpen [0]{}%
\providecommand \bibitemStop [0]{}%
\providecommand \bibitemNoStop [0]{.\EOS\space}%
\providecommand \EOS [0]{\spacefactor3000\relax}%
\providecommand \BibitemShut  [1]{\csname bibitem#1\endcsname}%
\let\auto@bib@innerbib\@empty
\bibitem [{\citenamefont {{Amaro-Seoane}}\ \emph {et~al.}(2017)\citenamefont
  {{Amaro-Seoane}}, \citenamefont {{Audley}}, \citenamefont {{Babak}},
  \citenamefont {{Baker}}, \citenamefont {{Barausse}}, \citenamefont
  {{Bender}}, \citenamefont {{Berti}}, \citenamefont {{Binetruy}},
  \citenamefont {{Born}}, \citenamefont {{Bortoluzzi}}, \citenamefont {{Camp}},
  \citenamefont {{Caprini}}, \citenamefont {{Cardoso}}, \citenamefont
  {{Colpi}}, \citenamefont {{Conklin}}, \citenamefont {{Cornish}},
  \citenamefont {{Cutler}}, \citenamefont {{Danzmann}}, \citenamefont
  {{Dolesi}}, \citenamefont {{Ferraioli}}, \citenamefont {{Ferroni}},
  \citenamefont {{Fitzsimons}}, \citenamefont {{Gair}}, \citenamefont {{Gesa
  Bote}}, \citenamefont {{Giardini}}, \citenamefont {{Gibert}}, \citenamefont
  {{Grimani}}, \citenamefont {{Halloin}}, \citenamefont {{Heinzel}},
  \citenamefont {{Hertog}}, \citenamefont {{Hewitson}}, \citenamefont
  {{Holley-Bockelmann}}, \citenamefont {{Hollington}}, \citenamefont
  {{Hueller}}, \citenamefont {{Inchauspe}}, \citenamefont {{Jetzer}},
  \citenamefont {{Karnesis}}, \citenamefont {{Killow}}, \citenamefont
  {{Klein}}, \citenamefont {{Klipstein}}, \citenamefont {{Korsakova}},
  \citenamefont {{Larson}}, \citenamefont {{Livas}}, \citenamefont {{Lloro}},
  \citenamefont {{Man}}, \citenamefont {{Mance}}, \citenamefont {{Martino}},
  \citenamefont {{Mateos}}, \citenamefont {{McKenzie}}, \citenamefont
  {{McWilliams}}, \citenamefont {{Miller}}, \citenamefont {{Mueller}},
  \citenamefont {{Nardini}}, \citenamefont {{Nelemans}}, \citenamefont
  {{Nofrarias}}, \citenamefont {{Petiteau}}, \citenamefont {{Pivato}},
  \citenamefont {{Plagnol}}, \citenamefont {{Porter}}, \citenamefont
  {{Reiche}}, \citenamefont {{Robertson}}, \citenamefont {{Robertson}},
  \citenamefont {{Rossi}}, \citenamefont {{Russano}}, \citenamefont {{Schutz}},
  \citenamefont {{Sesana}}, \citenamefont {{Shoemaker}}, \citenamefont
  {{Slutsky}}, \citenamefont {{Sopuerta}}, \citenamefont {{Sumner}},
  \citenamefont {{Tamanini}}, \citenamefont {{Thorpe}}, \citenamefont
  {{Troebs}}, \citenamefont {{Vallisneri}}, \citenamefont {{Vecchio}},
  \citenamefont {{Vetrugno}}, \citenamefont {{Vitale}}, \citenamefont
  {{Volonteri}}, \citenamefont {{Wanner}}, \citenamefont {{Ward}},
  \citenamefont {{Wass}}, \citenamefont {{Weber}}, \citenamefont {{Ziemer}},\
  and\ \citenamefont {{Zweifel}}}]{LISAproposal}%
  \BibitemOpen
  \bibfield  {author} {\bibinfo {author} {\bibfnamefont {P.}~\bibnamefont
  {{Amaro-Seoane}}}, \bibinfo {author} {\bibfnamefont {H.}~\bibnamefont
  {{Audley}}}, \bibinfo {author} {\bibfnamefont {S.}~\bibnamefont {{Babak}}},
  \bibinfo {author} {\bibfnamefont {J.}~\bibnamefont {{Baker}}}, \bibinfo
  {author} {\bibfnamefont {E.}~\bibnamefont {{Barausse}}}, \bibinfo {author}
  {\bibfnamefont {P.}~\bibnamefont {{Bender}}}, \bibinfo {author}
  {\bibfnamefont {E.}~\bibnamefont {{Berti}}}, \bibinfo {author} {\bibfnamefont
  {P.}~\bibnamefont {{Binetruy}}}, \bibinfo {author} {\bibfnamefont
  {M.}~\bibnamefont {{Born}}}, \bibinfo {author} {\bibfnamefont
  {D.}~\bibnamefont {{Bortoluzzi}}}, \bibinfo {author} {\bibfnamefont
  {J.}~\bibnamefont {{Camp}}}, \bibinfo {author} {\bibfnamefont
  {C.}~\bibnamefont {{Caprini}}}, \bibinfo {author} {\bibfnamefont
  {V.}~\bibnamefont {{Cardoso}}}, \bibinfo {author} {\bibfnamefont
  {M.}~\bibnamefont {{Colpi}}}, \bibinfo {author} {\bibfnamefont
  {J.}~\bibnamefont {{Conklin}}}, \bibinfo {author} {\bibfnamefont
  {N.}~\bibnamefont {{Cornish}}}, \bibinfo {author} {\bibfnamefont
  {C.}~\bibnamefont {{Cutler}}}, \bibinfo {author} {\bibfnamefont
  {K.}~\bibnamefont {{Danzmann}}}, \bibinfo {author} {\bibfnamefont
  {R.}~\bibnamefont {{Dolesi}}}, \bibinfo {author} {\bibfnamefont
  {L.}~\bibnamefont {{Ferraioli}}}, \bibinfo {author} {\bibfnamefont
  {V.}~\bibnamefont {{Ferroni}}}, \bibinfo {author} {\bibfnamefont
  {E.}~\bibnamefont {{Fitzsimons}}}, \bibinfo {author} {\bibfnamefont
  {J.}~\bibnamefont {{Gair}}}, \bibinfo {author} {\bibfnamefont
  {L.}~\bibnamefont {{Gesa Bote}}}, \bibinfo {author} {\bibfnamefont
  {D.}~\bibnamefont {{Giardini}}}, \bibinfo {author} {\bibfnamefont
  {F.}~\bibnamefont {{Gibert}}}, \bibinfo {author} {\bibfnamefont
  {C.}~\bibnamefont {{Grimani}}}, \bibinfo {author} {\bibfnamefont
  {H.}~\bibnamefont {{Halloin}}}, \bibinfo {author} {\bibfnamefont
  {G.}~\bibnamefont {{Heinzel}}}, \bibinfo {author} {\bibfnamefont
  {T.}~\bibnamefont {{Hertog}}}, \bibinfo {author} {\bibfnamefont
  {M.}~\bibnamefont {{Hewitson}}}, \bibinfo {author} {\bibfnamefont
  {K.}~\bibnamefont {{Holley-Bockelmann}}}, \bibinfo {author} {\bibfnamefont
  {D.}~\bibnamefont {{Hollington}}}, \bibinfo {author} {\bibfnamefont
  {M.}~\bibnamefont {{Hueller}}}, \bibinfo {author} {\bibfnamefont
  {H.}~\bibnamefont {{Inchauspe}}}, \bibinfo {author} {\bibfnamefont
  {P.}~\bibnamefont {{Jetzer}}}, \bibinfo {author} {\bibfnamefont
  {N.}~\bibnamefont {{Karnesis}}}, \bibinfo {author} {\bibfnamefont
  {C.}~\bibnamefont {{Killow}}}, \bibinfo {author} {\bibfnamefont
  {A.}~\bibnamefont {{Klein}}}, \bibinfo {author} {\bibfnamefont
  {B.}~\bibnamefont {{Klipstein}}}, \bibinfo {author} {\bibfnamefont
  {N.}~\bibnamefont {{Korsakova}}}, \bibinfo {author} {\bibfnamefont {S.~L.}\
  \bibnamefont {{Larson}}}, \bibinfo {author} {\bibfnamefont {J.}~\bibnamefont
  {{Livas}}}, \bibinfo {author} {\bibfnamefont {I.}~\bibnamefont {{Lloro}}},
  \bibinfo {author} {\bibfnamefont {N.}~\bibnamefont {{Man}}}, \bibinfo
  {author} {\bibfnamefont {D.}~\bibnamefont {{Mance}}}, \bibinfo {author}
  {\bibfnamefont {J.}~\bibnamefont {{Martino}}}, \bibinfo {author}
  {\bibfnamefont {I.}~\bibnamefont {{Mateos}}}, \bibinfo {author}
  {\bibfnamefont {K.}~\bibnamefont {{McKenzie}}}, \bibinfo {author}
  {\bibfnamefont {S.~T.}\ \bibnamefont {{McWilliams}}}, \bibinfo {author}
  {\bibfnamefont {C.}~\bibnamefont {{Miller}}}, \bibinfo {author}
  {\bibfnamefont {G.}~\bibnamefont {{Mueller}}}, \bibinfo {author}
  {\bibfnamefont {G.}~\bibnamefont {{Nardini}}}, \bibinfo {author}
  {\bibfnamefont {G.}~\bibnamefont {{Nelemans}}}, \bibinfo {author}
  {\bibfnamefont {M.}~\bibnamefont {{Nofrarias}}}, \bibinfo {author}
  {\bibfnamefont {A.}~\bibnamefont {{Petiteau}}}, \bibinfo {author}
  {\bibfnamefont {P.}~\bibnamefont {{Pivato}}}, \bibinfo {author}
  {\bibfnamefont {E.}~\bibnamefont {{Plagnol}}}, \bibinfo {author}
  {\bibfnamefont {E.}~\bibnamefont {{Porter}}}, \bibinfo {author}
  {\bibfnamefont {J.}~\bibnamefont {{Reiche}}}, \bibinfo {author}
  {\bibfnamefont {D.}~\bibnamefont {{Robertson}}}, \bibinfo {author}
  {\bibfnamefont {N.}~\bibnamefont {{Robertson}}}, \bibinfo {author}
  {\bibfnamefont {E.}~\bibnamefont {{Rossi}}}, \bibinfo {author} {\bibfnamefont
  {G.}~\bibnamefont {{Russano}}}, \bibinfo {author} {\bibfnamefont
  {B.}~\bibnamefont {{Schutz}}}, \bibinfo {author} {\bibfnamefont
  {A.}~\bibnamefont {{Sesana}}}, \bibinfo {author} {\bibfnamefont
  {D.}~\bibnamefont {{Shoemaker}}}, \bibinfo {author} {\bibfnamefont
  {J.}~\bibnamefont {{Slutsky}}}, \bibinfo {author} {\bibfnamefont {C.~F.}\
  \bibnamefont {{Sopuerta}}}, \bibinfo {author} {\bibfnamefont
  {T.}~\bibnamefont {{Sumner}}}, \bibinfo {author} {\bibfnamefont
  {N.}~\bibnamefont {{Tamanini}}}, \bibinfo {author} {\bibfnamefont
  {I.}~\bibnamefont {{Thorpe}}}, \bibinfo {author} {\bibfnamefont
  {M.}~\bibnamefont {{Troebs}}}, \bibinfo {author} {\bibfnamefont
  {M.}~\bibnamefont {{Vallisneri}}}, \bibinfo {author} {\bibfnamefont
  {A.}~\bibnamefont {{Vecchio}}}, \bibinfo {author} {\bibfnamefont
  {D.}~\bibnamefont {{Vetrugno}}}, \bibinfo {author} {\bibfnamefont
  {S.}~\bibnamefont {{Vitale}}}, \bibinfo {author} {\bibfnamefont
  {M.}~\bibnamefont {{Volonteri}}}, \bibinfo {author} {\bibfnamefont
  {G.}~\bibnamefont {{Wanner}}}, \bibinfo {author} {\bibfnamefont
  {H.}~\bibnamefont {{Ward}}}, \bibinfo {author} {\bibfnamefont
  {P.}~\bibnamefont {{Wass}}}, \bibinfo {author} {\bibfnamefont
  {W.}~\bibnamefont {{Weber}}}, \bibinfo {author} {\bibfnamefont
  {J.}~\bibnamefont {{Ziemer}}}, \ and\ \bibinfo {author} {\bibfnamefont
  {P.}~\bibnamefont {{Zweifel}}},\ }\href@noop {} {\bibfield  {journal}
  {\bibinfo  {journal} {arXiv e-prints}\ ,\ \bibinfo {eid} {arXiv:1702.00786}}
  (\bibinfo {year} {2017})},\ \Eprint {http://arxiv.org/abs/1702.00786}
  {arXiv:1702.00786 [astro-ph.IM]} \BibitemShut {NoStop}%
\bibitem [{\citenamefont {{Luo}}\ \emph {et~al.}(2016)\citenamefont {{Luo}},
  \citenamefont {{Chen}}, \citenamefont {{Duan}}, \citenamefont {{Gong}},
  \citenamefont {{Hu}}, \citenamefont {{Ji}}, \citenamefont {{Liu}},
  \citenamefont {{Mei}}, \citenamefont {{Milyukov}}, \citenamefont {{Sazhin}},
  \citenamefont {{Shao}}, \citenamefont {{Toth}}, \citenamefont {{Tu}},
  \citenamefont {{Wang}}, \citenamefont {{Wang}}, \citenamefont {{Yeh}},
  \citenamefont {{Zhan}}, \citenamefont {{Zhang}}, \citenamefont {{Zharov}},\
  and\ \citenamefont {{Zhou}}}]{TianQin}%
  \BibitemOpen
  \bibfield  {author} {\bibinfo {author} {\bibfnamefont {J.}~\bibnamefont
  {{Luo}}}, \bibinfo {author} {\bibfnamefont {L.-S.}\ \bibnamefont {{Chen}}},
  \bibinfo {author} {\bibfnamefont {H.-Z.}\ \bibnamefont {{Duan}}}, \bibinfo
  {author} {\bibfnamefont {Y.-G.}\ \bibnamefont {{Gong}}}, \bibinfo {author}
  {\bibfnamefont {S.}~\bibnamefont {{Hu}}}, \bibinfo {author} {\bibfnamefont
  {J.}~\bibnamefont {{Ji}}}, \bibinfo {author} {\bibfnamefont {Q.}~\bibnamefont
  {{Liu}}}, \bibinfo {author} {\bibfnamefont {J.}~\bibnamefont {{Mei}}},
  \bibinfo {author} {\bibfnamefont {V.}~\bibnamefont {{Milyukov}}}, \bibinfo
  {author} {\bibfnamefont {M.}~\bibnamefont {{Sazhin}}}, \bibinfo {author}
  {\bibfnamefont {C.-G.}\ \bibnamefont {{Shao}}}, \bibinfo {author}
  {\bibfnamefont {V.~T.}\ \bibnamefont {{Toth}}}, \bibinfo {author}
  {\bibfnamefont {H.-B.}\ \bibnamefont {{Tu}}}, \bibinfo {author}
  {\bibfnamefont {Y.}~\bibnamefont {{Wang}}}, \bibinfo {author} {\bibfnamefont
  {Y.}~\bibnamefont {{Wang}}}, \bibinfo {author} {\bibfnamefont {H.-C.}\
  \bibnamefont {{Yeh}}}, \bibinfo {author} {\bibfnamefont {M.-S.}\ \bibnamefont
  {{Zhan}}}, \bibinfo {author} {\bibfnamefont {Y.}~\bibnamefont {{Zhang}}},
  \bibinfo {author} {\bibfnamefont {V.}~\bibnamefont {{Zharov}}}, \ and\
  \bibinfo {author} {\bibfnamefont {Z.-B.}\ \bibnamefont {{Zhou}}},\ }\href
  {\doibase 10.1088/0264-9381/33/3/035010} {\bibfield  {journal} {\bibinfo
  {journal} {Classical and Quantum Gravity}\ }\textbf {\bibinfo {volume}
  {33}},\ \bibinfo {eid} {035010} (\bibinfo {year} {2016})},\ \Eprint
  {http://arxiv.org/abs/1512.02076} {arXiv:1512.02076 [astro-ph.IM]}
  \BibitemShut {NoStop}%
\bibitem [{Note1()}]{Note1}%
  \BibitemOpen
  \bibinfo {note} {The term ``space weather'' is often focused on the local
  radiation environment of the Earth, but is used through out this paper in its
  more general sense, describing the time-varying interplanetary environment
  driven by changes in the solar wind and solar output.}\BibitemShut {Stop}%
\bibitem [{\citenamefont {{Armano}}\ \emph
  {et~al.}(2017{\natexlab{a}})\citenamefont {{Armano}}, \citenamefont
  {{Audley}}, \citenamefont {{Auger}}, \citenamefont {{Baird}}, \citenamefont
  {{Binetruy}}, \citenamefont {{Born}}, \citenamefont {{Bortoluzzi}},
  \citenamefont {{Brandt}}, \citenamefont {{Bursi}}, \citenamefont {{Caleno}},
  \citenamefont {{Cavalleri}}, \citenamefont {{Cesarini}}, \citenamefont
  {{Cruise}}, \citenamefont {{Danzmann}}, \citenamefont {{de Deus Silva}},
  \citenamefont {{Diepholz}}, \citenamefont {{Dolesi}}, \citenamefont
  {{Dunbar}}, \citenamefont {{Ferraioli}}, \citenamefont {{Ferroni}},
  \citenamefont {{Fitzsimons}}, \citenamefont {{Flatscher}}, \citenamefont
  {{Freschi}}, \citenamefont {{Gallegos}}, \citenamefont {{Garc{\'\i}a
  Marirrodriga}}, \citenamefont {{Gerndt}}, \citenamefont {{Gesa}},
  \citenamefont {{Gibert}}, \citenamefont {{Giardini}}, \citenamefont
  {{Giusteri}}, \citenamefont {{Grimani}}, \citenamefont {{Grzymisch}},
  \citenamefont {{Harrison}}, \citenamefont {{Heinzel}}, \citenamefont
  {{Hewitson}}, \citenamefont {{Hollington}}, \citenamefont {{Hueller}},
  \citenamefont {{Huesler}}, \citenamefont {{Inchausp{\'e}}}, \citenamefont
  {{Jennrich}}, \citenamefont {{Jetzer}}, \citenamefont {{Johland er}},
  \citenamefont {{Karnesis}}, \citenamefont {{Kaune}}, \citenamefont
  {{Killow}}, \citenamefont {{Korsakova}}, \citenamefont {{Lloro}},
  \citenamefont {{Liu}}, \citenamefont {{L{\'o}pez-Zaragoza}}, \citenamefont
  {{Maarschalkerweerd}}, \citenamefont {{Madden}}, \citenamefont {{Mance}},
  \citenamefont {{Mart{\'\i}n}}, \citenamefont {{Martin-Polo}}, \citenamefont
  {{Martino}}, \citenamefont {{Martin-Porqueras}}, \citenamefont {{Mateos}},
  \citenamefont {{McNamara}}, \citenamefont {{Mendes}}, \citenamefont
  {{Mendes}}, \citenamefont {{Moroni}}, \citenamefont {{Nofrarias}},
  \citenamefont {{Paczkowski}}, \citenamefont {{Perreur-Lloyd}}, \citenamefont
  {{Petiteau}}, \citenamefont {{Pivato}}, \citenamefont {{Plagnol}},
  \citenamefont {{Prat}}, \citenamefont {{Ragnit}}, \citenamefont
  {{Ramos-Castro}}, \citenamefont {{Reiche}}, \citenamefont {{Romera Perez}},
  \citenamefont {{Robertson}}, \citenamefont {{Rozemeijer}}, \citenamefont
  {{Rivas}}, \citenamefont {{Russano}}, \citenamefont {{Sarra}}, \citenamefont
  {{Schleicher}}, \citenamefont {{Slutsky}}, \citenamefont {{Sopuerta}},
  \citenamefont {{Sumner}}, \citenamefont {{Texier}}, \citenamefont {{Thorpe}},
  \citenamefont {{Trenkel}}, \citenamefont {{Vetrugno}}, \citenamefont
  {{Vitale}}, \citenamefont {{Wanner}}, \citenamefont {{Ward}}, \citenamefont
  {{Wass}}, \citenamefont {{Wealthy}}, \citenamefont {{Weber}}, \citenamefont
  {{Wittchen}}, \citenamefont {{Zanoni}}, \citenamefont {{Ziegler}},
  \citenamefont {{Zweifel}},\ and\ \citenamefont {{LISA Pathfinder
  Collaboration}}}]{PathfinderCharge}%
  \BibitemOpen
  \bibfield  {author} {\bibinfo {author} {\bibfnamefont {M.}~\bibnamefont
  {{Armano}}}, \bibinfo {author} {\bibfnamefont {H.}~\bibnamefont {{Audley}}},
  \bibinfo {author} {\bibfnamefont {G.}~\bibnamefont {{Auger}}}, \bibinfo
  {author} {\bibfnamefont {J.~T.}\ \bibnamefont {{Baird}}}, \bibinfo {author}
  {\bibfnamefont {P.}~\bibnamefont {{Binetruy}}}, \bibinfo {author}
  {\bibfnamefont {M.}~\bibnamefont {{Born}}}, \bibinfo {author} {\bibfnamefont
  {D.}~\bibnamefont {{Bortoluzzi}}}, \bibinfo {author} {\bibfnamefont
  {N.}~\bibnamefont {{Brandt}}}, \bibinfo {author} {\bibfnamefont
  {A.}~\bibnamefont {{Bursi}}}, \bibinfo {author} {\bibfnamefont
  {M.}~\bibnamefont {{Caleno}}}, \bibinfo {author} {\bibfnamefont
  {A.}~\bibnamefont {{Cavalleri}}}, \bibinfo {author} {\bibfnamefont
  {A.}~\bibnamefont {{Cesarini}}}, \bibinfo {author} {\bibfnamefont
  {M.}~\bibnamefont {{Cruise}}}, \bibinfo {author} {\bibfnamefont
  {K.}~\bibnamefont {{Danzmann}}}, \bibinfo {author} {\bibfnamefont
  {M.}~\bibnamefont {{de Deus Silva}}}, \bibinfo {author} {\bibfnamefont
  {I.}~\bibnamefont {{Diepholz}}}, \bibinfo {author} {\bibfnamefont
  {R.}~\bibnamefont {{Dolesi}}}, \bibinfo {author} {\bibfnamefont
  {N.}~\bibnamefont {{Dunbar}}}, \bibinfo {author} {\bibfnamefont
  {L.}~\bibnamefont {{Ferraioli}}}, \bibinfo {author} {\bibfnamefont
  {V.}~\bibnamefont {{Ferroni}}}, \bibinfo {author} {\bibfnamefont {E.~D.}\
  \bibnamefont {{Fitzsimons}}}, \bibinfo {author} {\bibfnamefont
  {R.}~\bibnamefont {{Flatscher}}}, \bibinfo {author} {\bibfnamefont
  {M.}~\bibnamefont {{Freschi}}}, \bibinfo {author} {\bibfnamefont
  {J.}~\bibnamefont {{Gallegos}}}, \bibinfo {author} {\bibfnamefont
  {C.}~\bibnamefont {{Garc{\'\i}a Marirrodriga}}}, \bibinfo {author}
  {\bibfnamefont {R.}~\bibnamefont {{Gerndt}}}, \bibinfo {author}
  {\bibfnamefont {L.}~\bibnamefont {{Gesa}}}, \bibinfo {author} {\bibfnamefont
  {F.}~\bibnamefont {{Gibert}}}, \bibinfo {author} {\bibfnamefont
  {D.}~\bibnamefont {{Giardini}}}, \bibinfo {author} {\bibfnamefont
  {R.}~\bibnamefont {{Giusteri}}}, \bibinfo {author} {\bibfnamefont
  {C.}~\bibnamefont {{Grimani}}}, \bibinfo {author} {\bibfnamefont
  {J.}~\bibnamefont {{Grzymisch}}}, \bibinfo {author} {\bibfnamefont
  {I.}~\bibnamefont {{Harrison}}}, \bibinfo {author} {\bibfnamefont
  {G.}~\bibnamefont {{Heinzel}}}, \bibinfo {author} {\bibfnamefont
  {M.}~\bibnamefont {{Hewitson}}}, \bibinfo {author} {\bibfnamefont
  {D.}~\bibnamefont {{Hollington}}}, \bibinfo {author} {\bibfnamefont
  {M.}~\bibnamefont {{Hueller}}}, \bibinfo {author} {\bibfnamefont
  {J.}~\bibnamefont {{Huesler}}}, \bibinfo {author} {\bibfnamefont
  {H.}~\bibnamefont {{Inchausp{\'e}}}}, \bibinfo {author} {\bibfnamefont
  {O.}~\bibnamefont {{Jennrich}}}, \bibinfo {author} {\bibfnamefont
  {P.}~\bibnamefont {{Jetzer}}}, \bibinfo {author} {\bibfnamefont
  {B.}~\bibnamefont {{Johland er}}}, \bibinfo {author} {\bibfnamefont
  {N.}~\bibnamefont {{Karnesis}}}, \bibinfo {author} {\bibfnamefont
  {B.}~\bibnamefont {{Kaune}}}, \bibinfo {author} {\bibfnamefont {C.~J.}\
  \bibnamefont {{Killow}}}, \bibinfo {author} {\bibfnamefont {N.}~\bibnamefont
  {{Korsakova}}}, \bibinfo {author} {\bibfnamefont {I.}~\bibnamefont
  {{Lloro}}}, \bibinfo {author} {\bibfnamefont {L.}~\bibnamefont {{Liu}}},
  \bibinfo {author} {\bibfnamefont {J.~P.}\ \bibnamefont
  {{L{\'o}pez-Zaragoza}}}, \bibinfo {author} {\bibfnamefont {R.}~\bibnamefont
  {{Maarschalkerweerd}}}, \bibinfo {author} {\bibfnamefont {S.}~\bibnamefont
  {{Madden}}}, \bibinfo {author} {\bibfnamefont {D.}~\bibnamefont {{Mance}}},
  \bibinfo {author} {\bibfnamefont {V.}~\bibnamefont {{Mart{\'\i}n}}}, \bibinfo
  {author} {\bibfnamefont {L.}~\bibnamefont {{Martin-Polo}}}, \bibinfo {author}
  {\bibfnamefont {J.}~\bibnamefont {{Martino}}}, \bibinfo {author}
  {\bibfnamefont {F.}~\bibnamefont {{Martin-Porqueras}}}, \bibinfo {author}
  {\bibfnamefont {I.}~\bibnamefont {{Mateos}}}, \bibinfo {author}
  {\bibfnamefont {P.~W.}\ \bibnamefont {{McNamara}}}, \bibinfo {author}
  {\bibfnamefont {J.}~\bibnamefont {{Mendes}}}, \bibinfo {author}
  {\bibfnamefont {L.}~\bibnamefont {{Mendes}}}, \bibinfo {author}
  {\bibfnamefont {A.}~\bibnamefont {{Moroni}}}, \bibinfo {author}
  {\bibfnamefont {M.}~\bibnamefont {{Nofrarias}}}, \bibinfo {author}
  {\bibfnamefont {S.}~\bibnamefont {{Paczkowski}}}, \bibinfo {author}
  {\bibfnamefont {M.}~\bibnamefont {{Perreur-Lloyd}}}, \bibinfo {author}
  {\bibfnamefont {A.}~\bibnamefont {{Petiteau}}}, \bibinfo {author}
  {\bibfnamefont {P.}~\bibnamefont {{Pivato}}}, \bibinfo {author}
  {\bibfnamefont {E.}~\bibnamefont {{Plagnol}}}, \bibinfo {author}
  {\bibfnamefont {P.}~\bibnamefont {{Prat}}}, \bibinfo {author} {\bibfnamefont
  {U.}~\bibnamefont {{Ragnit}}}, \bibinfo {author} {\bibfnamefont
  {J.}~\bibnamefont {{Ramos-Castro}}}, \bibinfo {author} {\bibfnamefont
  {J.}~\bibnamefont {{Reiche}}}, \bibinfo {author} {\bibfnamefont {J.~A.}\
  \bibnamefont {{Romera Perez}}}, \bibinfo {author} {\bibfnamefont {D.~I.}\
  \bibnamefont {{Robertson}}}, \bibinfo {author} {\bibfnamefont
  {H.}~\bibnamefont {{Rozemeijer}}}, \bibinfo {author} {\bibfnamefont
  {F.}~\bibnamefont {{Rivas}}}, \bibinfo {author} {\bibfnamefont
  {G.}~\bibnamefont {{Russano}}}, \bibinfo {author} {\bibfnamefont
  {P.}~\bibnamefont {{Sarra}}}, \bibinfo {author} {\bibfnamefont
  {A.}~\bibnamefont {{Schleicher}}}, \bibinfo {author} {\bibfnamefont
  {J.}~\bibnamefont {{Slutsky}}}, \bibinfo {author} {\bibfnamefont
  {C.}~\bibnamefont {{Sopuerta}}}, \bibinfo {author} {\bibfnamefont {T.~J.}\
  \bibnamefont {{Sumner}}}, \bibinfo {author} {\bibfnamefont {D.}~\bibnamefont
  {{Texier}}}, \bibinfo {author} {\bibfnamefont {J.~I.}\ \bibnamefont
  {{Thorpe}}}, \bibinfo {author} {\bibfnamefont {C.}~\bibnamefont {{Trenkel}}},
  \bibinfo {author} {\bibfnamefont {D.}~\bibnamefont {{Vetrugno}}}, \bibinfo
  {author} {\bibfnamefont {S.}~\bibnamefont {{Vitale}}}, \bibinfo {author}
  {\bibfnamefont {G.}~\bibnamefont {{Wanner}}}, \bibinfo {author}
  {\bibfnamefont {H.}~\bibnamefont {{Ward}}}, \bibinfo {author} {\bibfnamefont
  {P.~J.}\ \bibnamefont {{Wass}}}, \bibinfo {author} {\bibfnamefont
  {D.}~\bibnamefont {{Wealthy}}}, \bibinfo {author} {\bibfnamefont {W.~J.}\
  \bibnamefont {{Weber}}}, \bibinfo {author} {\bibfnamefont {A.}~\bibnamefont
  {{Wittchen}}}, \bibinfo {author} {\bibfnamefont {C.}~\bibnamefont
  {{Zanoni}}}, \bibinfo {author} {\bibfnamefont {T.}~\bibnamefont {{Ziegler}}},
  \bibinfo {author} {\bibfnamefont {P.}~\bibnamefont {{Zweifel}}}, \ and\
  \bibinfo {author} {\bibnamefont {{LISA Pathfinder Collaboration}}},\ }\href
  {\doibase 10.1103/PhysRevLett.118.171101} {\bibfield  {journal} {\bibinfo
  {journal} {\prl}\ }\textbf {\bibinfo {volume} {118}},\ \bibinfo {eid}
  {171101} (\bibinfo {year} {2017}{\natexlab{a}})},\ \Eprint
  {http://arxiv.org/abs/1702.04633} {arXiv:1702.04633 [astro-ph.IM]}
  \BibitemShut {NoStop}%
\bibitem [{\citenamefont {{Armano}}\ \emph
  {et~al.}(2017{\natexlab{b}})\citenamefont {{Armano}}, \citenamefont
  {{Audley}}, \citenamefont {{Auger}}, \citenamefont {{Baird}}, \citenamefont
  {{Binetruy}}, \citenamefont {{Born}}, \citenamefont {{Bortoluzzi}},
  \citenamefont {{Brandt}}, \citenamefont {{Bursi}}, \citenamefont {{Caleno}},
  \citenamefont {{Cavalleri}}, \citenamefont {{Cesarini}}, \citenamefont
  {{Cruise}}, \citenamefont {{Danzmann}}, \citenamefont {{de Deus Silva}},
  \citenamefont {{Diepholz}}, \citenamefont {{Dolesi}}, \citenamefont
  {{Dunbar}}, \citenamefont {{Ferraioli}}, \citenamefont {{Ferroni}},
  \citenamefont {{Fitzsimons}}, \citenamefont {{Flatscher}}, \citenamefont
  {{Freschi}}, \citenamefont {{Gallegos}}, \citenamefont {{Garc{\'\i}a
  Marirrodriga}}, \citenamefont {{Gerndt}}, \citenamefont {{Gesa}},
  \citenamefont {{Gibert}}, \citenamefont {{Giardini}}, \citenamefont
  {{Giusteri}}, \citenamefont {{Grimani}}, \citenamefont {{Grzymisch}},
  \citenamefont {{Harrison}}, \citenamefont {{Heinzel}}, \citenamefont
  {{Hewitson}}, \citenamefont {{Hollington}}, \citenamefont {{Hueller}},
  \citenamefont {{Huesler}}, \citenamefont {{Inchausp{\'e}}}, \citenamefont
  {{Jennrich}}, \citenamefont {{Jetzer}}, \citenamefont {{Johland er}},
  \citenamefont {{Karnesis}}, \citenamefont {{Kaune}}, \citenamefont
  {{Killow}}, \citenamefont {{Korsakova}}, \citenamefont {{Lloro}},
  \citenamefont {{Liu}}, \citenamefont {{L{\'o}pez-Zaragoza}}, \citenamefont
  {{Maarschalkerweerd}}, \citenamefont {{Madden}}, \citenamefont {{Mance}},
  \citenamefont {{Mart{\'\i}n}}, \citenamefont {{Martin-Polo}}, \citenamefont
  {{Martino}}, \citenamefont {{Martin-Porqueras}}, \citenamefont {{Mateos}},
  \citenamefont {{McNamara}}, \citenamefont {{Mendes}}, \citenamefont
  {{Mendes}}, \citenamefont {{Moroni}}, \citenamefont {{Nofrarias}},
  \citenamefont {{Paczkowski}}, \citenamefont {{Perreur-Lloyd}}, \citenamefont
  {{Petiteau}}, \citenamefont {{Pivato}}, \citenamefont {{Plagnol}},
  \citenamefont {{Prat}}, \citenamefont {{Ragnit}}, \citenamefont
  {{Ramos-Castro}}, \citenamefont {{Reiche}}, \citenamefont {{Romera Perez}},
  \citenamefont {{Robertson}}, \citenamefont {{Rozemeijer}}, \citenamefont
  {{Rivas}}, \citenamefont {{Russano}}, \citenamefont {{Sarra}}, \citenamefont
  {{Schleicher}}, \citenamefont {{Slutsky}}, \citenamefont {{Sopuerta}},
  \citenamefont {{Sumner}}, \citenamefont {{Texier}}, \citenamefont {{Thorpe}},
  \citenamefont {{Trenkel}}, \citenamefont {{Vetrugno}}, \citenamefont
  {{Vitale}}, \citenamefont {{Wanner}}, \citenamefont {{Ward}}, \citenamefont
  {{Wass}}, \citenamefont {{Wealthy}}, \citenamefont {{Weber}}, \citenamefont
  {{Wittchen}}, \citenamefont {{Zanoni}}, \citenamefont {{Ziegler}},
  \citenamefont {{Zweifel}},\ and\ \citenamefont {{LISA Pathfinder
  Collaboration}}}]{PathfinderAccel}%
  \BibitemOpen
  \bibfield  {author} {\bibinfo {author} {\bibfnamefont {M.}~\bibnamefont
  {{Armano}}}, \bibinfo {author} {\bibfnamefont {H.}~\bibnamefont {{Audley}}},
  \bibinfo {author} {\bibfnamefont {G.}~\bibnamefont {{Auger}}}, \bibinfo
  {author} {\bibfnamefont {J.~T.}\ \bibnamefont {{Baird}}}, \bibinfo {author}
  {\bibfnamefont {P.}~\bibnamefont {{Binetruy}}}, \bibinfo {author}
  {\bibfnamefont {M.}~\bibnamefont {{Born}}}, \bibinfo {author} {\bibfnamefont
  {D.}~\bibnamefont {{Bortoluzzi}}}, \bibinfo {author} {\bibfnamefont
  {N.}~\bibnamefont {{Brandt}}}, \bibinfo {author} {\bibfnamefont
  {A.}~\bibnamefont {{Bursi}}}, \bibinfo {author} {\bibfnamefont
  {M.}~\bibnamefont {{Caleno}}}, \bibinfo {author} {\bibfnamefont
  {A.}~\bibnamefont {{Cavalleri}}}, \bibinfo {author} {\bibfnamefont
  {A.}~\bibnamefont {{Cesarini}}}, \bibinfo {author} {\bibfnamefont
  {M.}~\bibnamefont {{Cruise}}}, \bibinfo {author} {\bibfnamefont
  {K.}~\bibnamefont {{Danzmann}}}, \bibinfo {author} {\bibfnamefont
  {M.}~\bibnamefont {{de Deus Silva}}}, \bibinfo {author} {\bibfnamefont
  {I.}~\bibnamefont {{Diepholz}}}, \bibinfo {author} {\bibfnamefont
  {R.}~\bibnamefont {{Dolesi}}}, \bibinfo {author} {\bibfnamefont
  {N.}~\bibnamefont {{Dunbar}}}, \bibinfo {author} {\bibfnamefont
  {L.}~\bibnamefont {{Ferraioli}}}, \bibinfo {author} {\bibfnamefont
  {V.}~\bibnamefont {{Ferroni}}}, \bibinfo {author} {\bibfnamefont {E.~D.}\
  \bibnamefont {{Fitzsimons}}}, \bibinfo {author} {\bibfnamefont
  {R.}~\bibnamefont {{Flatscher}}}, \bibinfo {author} {\bibfnamefont
  {M.}~\bibnamefont {{Freschi}}}, \bibinfo {author} {\bibfnamefont
  {J.}~\bibnamefont {{Gallegos}}}, \bibinfo {author} {\bibfnamefont
  {C.}~\bibnamefont {{Garc{\'\i}a Marirrodriga}}}, \bibinfo {author}
  {\bibfnamefont {R.}~\bibnamefont {{Gerndt}}}, \bibinfo {author}
  {\bibfnamefont {L.}~\bibnamefont {{Gesa}}}, \bibinfo {author} {\bibfnamefont
  {F.}~\bibnamefont {{Gibert}}}, \bibinfo {author} {\bibfnamefont
  {D.}~\bibnamefont {{Giardini}}}, \bibinfo {author} {\bibfnamefont
  {R.}~\bibnamefont {{Giusteri}}}, \bibinfo {author} {\bibfnamefont
  {C.}~\bibnamefont {{Grimani}}}, \bibinfo {author} {\bibfnamefont
  {J.}~\bibnamefont {{Grzymisch}}}, \bibinfo {author} {\bibfnamefont
  {I.}~\bibnamefont {{Harrison}}}, \bibinfo {author} {\bibfnamefont
  {G.}~\bibnamefont {{Heinzel}}}, \bibinfo {author} {\bibfnamefont
  {M.}~\bibnamefont {{Hewitson}}}, \bibinfo {author} {\bibfnamefont
  {D.}~\bibnamefont {{Hollington}}}, \bibinfo {author} {\bibfnamefont
  {M.}~\bibnamefont {{Hueller}}}, \bibinfo {author} {\bibfnamefont
  {J.}~\bibnamefont {{Huesler}}}, \bibinfo {author} {\bibfnamefont
  {H.}~\bibnamefont {{Inchausp{\'e}}}}, \bibinfo {author} {\bibfnamefont
  {O.}~\bibnamefont {{Jennrich}}}, \bibinfo {author} {\bibfnamefont
  {P.}~\bibnamefont {{Jetzer}}}, \bibinfo {author} {\bibfnamefont
  {B.}~\bibnamefont {{Johland er}}}, \bibinfo {author} {\bibfnamefont
  {N.}~\bibnamefont {{Karnesis}}}, \bibinfo {author} {\bibfnamefont
  {B.}~\bibnamefont {{Kaune}}}, \bibinfo {author} {\bibfnamefont {C.~J.}\
  \bibnamefont {{Killow}}}, \bibinfo {author} {\bibfnamefont {N.}~\bibnamefont
  {{Korsakova}}}, \bibinfo {author} {\bibfnamefont {I.}~\bibnamefont
  {{Lloro}}}, \bibinfo {author} {\bibfnamefont {L.}~\bibnamefont {{Liu}}},
  \bibinfo {author} {\bibfnamefont {J.~P.}\ \bibnamefont
  {{L{\'o}pez-Zaragoza}}}, \bibinfo {author} {\bibfnamefont {R.}~\bibnamefont
  {{Maarschalkerweerd}}}, \bibinfo {author} {\bibfnamefont {S.}~\bibnamefont
  {{Madden}}}, \bibinfo {author} {\bibfnamefont {D.}~\bibnamefont {{Mance}}},
  \bibinfo {author} {\bibfnamefont {V.}~\bibnamefont {{Mart{\'\i}n}}}, \bibinfo
  {author} {\bibfnamefont {L.}~\bibnamefont {{Martin-Polo}}}, \bibinfo {author}
  {\bibfnamefont {J.}~\bibnamefont {{Martino}}}, \bibinfo {author}
  {\bibfnamefont {F.}~\bibnamefont {{Martin-Porqueras}}}, \bibinfo {author}
  {\bibfnamefont {I.}~\bibnamefont {{Mateos}}}, \bibinfo {author}
  {\bibfnamefont {P.~W.}\ \bibnamefont {{McNamara}}}, \bibinfo {author}
  {\bibfnamefont {J.}~\bibnamefont {{Mendes}}}, \bibinfo {author}
  {\bibfnamefont {L.}~\bibnamefont {{Mendes}}}, \bibinfo {author}
  {\bibfnamefont {A.}~\bibnamefont {{Moroni}}}, \bibinfo {author}
  {\bibfnamefont {M.}~\bibnamefont {{Nofrarias}}}, \bibinfo {author}
  {\bibfnamefont {S.}~\bibnamefont {{Paczkowski}}}, \bibinfo {author}
  {\bibfnamefont {M.}~\bibnamefont {{Perreur-Lloyd}}}, \bibinfo {author}
  {\bibfnamefont {A.}~\bibnamefont {{Petiteau}}}, \bibinfo {author}
  {\bibfnamefont {P.}~\bibnamefont {{Pivato}}}, \bibinfo {author}
  {\bibfnamefont {E.}~\bibnamefont {{Plagnol}}}, \bibinfo {author}
  {\bibfnamefont {P.}~\bibnamefont {{Prat}}}, \bibinfo {author} {\bibfnamefont
  {U.}~\bibnamefont {{Ragnit}}}, \bibinfo {author} {\bibfnamefont
  {J.}~\bibnamefont {{Ramos-Castro}}}, \bibinfo {author} {\bibfnamefont
  {J.}~\bibnamefont {{Reiche}}}, \bibinfo {author} {\bibfnamefont {J.~A.}\
  \bibnamefont {{Romera Perez}}}, \bibinfo {author} {\bibfnamefont {D.~I.}\
  \bibnamefont {{Robertson}}}, \bibinfo {author} {\bibfnamefont
  {H.}~\bibnamefont {{Rozemeijer}}}, \bibinfo {author} {\bibfnamefont
  {F.}~\bibnamefont {{Rivas}}}, \bibinfo {author} {\bibfnamefont
  {G.}~\bibnamefont {{Russano}}}, \bibinfo {author} {\bibfnamefont
  {P.}~\bibnamefont {{Sarra}}}, \bibinfo {author} {\bibfnamefont
  {A.}~\bibnamefont {{Schleicher}}}, \bibinfo {author} {\bibfnamefont
  {J.}~\bibnamefont {{Slutsky}}}, \bibinfo {author} {\bibfnamefont
  {C.}~\bibnamefont {{Sopuerta}}}, \bibinfo {author} {\bibfnamefont {T.~J.}\
  \bibnamefont {{Sumner}}}, \bibinfo {author} {\bibfnamefont {D.}~\bibnamefont
  {{Texier}}}, \bibinfo {author} {\bibfnamefont {J.~I.}\ \bibnamefont
  {{Thorpe}}}, \bibinfo {author} {\bibfnamefont {C.}~\bibnamefont {{Trenkel}}},
  \bibinfo {author} {\bibfnamefont {D.}~\bibnamefont {{Vetrugno}}}, \bibinfo
  {author} {\bibfnamefont {S.}~\bibnamefont {{Vitale}}}, \bibinfo {author}
  {\bibfnamefont {G.}~\bibnamefont {{Wanner}}}, \bibinfo {author}
  {\bibfnamefont {H.}~\bibnamefont {{Ward}}}, \bibinfo {author} {\bibfnamefont
  {P.~J.}\ \bibnamefont {{Wass}}}, \bibinfo {author} {\bibfnamefont
  {D.}~\bibnamefont {{Wealthy}}}, \bibinfo {author} {\bibfnamefont {W.~J.}\
  \bibnamefont {{Weber}}}, \bibinfo {author} {\bibfnamefont {A.}~\bibnamefont
  {{Wittchen}}}, \bibinfo {author} {\bibfnamefont {C.}~\bibnamefont
  {{Zanoni}}}, \bibinfo {author} {\bibfnamefont {T.}~\bibnamefont {{Ziegler}}},
  \bibinfo {author} {\bibfnamefont {P.}~\bibnamefont {{Zweifel}}}, \ and\
  \bibinfo {author} {\bibnamefont {{LISA Pathfinder Collaboration}}},\ }\href
  {\doibase 10.1103/PhysRevLett.118.171101} {\bibfield  {journal} {\bibinfo
  {journal} {\prl}\ }\textbf {\bibinfo {volume} {118}},\ \bibinfo {eid}
  {171101} (\bibinfo {year} {2017}{\natexlab{b}})},\ \Eprint
  {http://arxiv.org/abs/1702.04633} {arXiv:1702.04633 [astro-ph.IM]}
  \BibitemShut {NoStop}%
\bibitem [{\citenamefont {{Fr{\"o}hlich}}\ \emph {et~al.}(1995)\citenamefont
  {{Fr{\"o}hlich}}, \citenamefont {{Romero}}, \citenamefont {{Roth}},
  \citenamefont {{Wehrli}}, \citenamefont {{Andersen}}, \citenamefont
  {{Appourchaux}}, \citenamefont {{Domingo}}, \citenamefont {{Telljohann}},
  \citenamefont {{Berthomieu}}, \citenamefont {{Delache}}, \citenamefont
  {{Provost}}, \citenamefont {{Toutain}}, \citenamefont {{Crommelynck}},
  \citenamefont {{Chevalier}}, \citenamefont {{Fichot}}, \citenamefont
  {{D{\"a}ppen}}, \citenamefont {{Gough}}, \citenamefont {{Hoeksema}},
  \citenamefont {{Jim{\'e}nez}}, \citenamefont {{G{\'o}mez}}, \citenamefont
  {{Herreros}}, \citenamefont {{Cort{\'e}s}}, \citenamefont {{Jones}},
  \citenamefont {{Pap}},\ and\ \citenamefont {{Willson}}}]{VIRGO1}%
  \BibitemOpen
  \bibfield  {author} {\bibinfo {author} {\bibfnamefont {C.}~\bibnamefont
  {{Fr{\"o}hlich}}}, \bibinfo {author} {\bibfnamefont {J.}~\bibnamefont
  {{Romero}}}, \bibinfo {author} {\bibfnamefont {H.}~\bibnamefont {{Roth}}},
  \bibinfo {author} {\bibfnamefont {C.}~\bibnamefont {{Wehrli}}}, \bibinfo
  {author} {\bibfnamefont {B.~N.}\ \bibnamefont {{Andersen}}}, \bibinfo
  {author} {\bibfnamefont {T.}~\bibnamefont {{Appourchaux}}}, \bibinfo {author}
  {\bibfnamefont {V.}~\bibnamefont {{Domingo}}}, \bibinfo {author}
  {\bibfnamefont {U.}~\bibnamefont {{Telljohann}}}, \bibinfo {author}
  {\bibfnamefont {G.}~\bibnamefont {{Berthomieu}}}, \bibinfo {author}
  {\bibfnamefont {P.}~\bibnamefont {{Delache}}}, \bibinfo {author}
  {\bibfnamefont {J.}~\bibnamefont {{Provost}}}, \bibinfo {author}
  {\bibfnamefont {T.}~\bibnamefont {{Toutain}}}, \bibinfo {author}
  {\bibfnamefont {D.~A.}\ \bibnamefont {{Crommelynck}}}, \bibinfo {author}
  {\bibfnamefont {A.}~\bibnamefont {{Chevalier}}}, \bibinfo {author}
  {\bibfnamefont {A.}~\bibnamefont {{Fichot}}}, \bibinfo {author}
  {\bibfnamefont {W.}~\bibnamefont {{D{\"a}ppen}}}, \bibinfo {author}
  {\bibfnamefont {D.}~\bibnamefont {{Gough}}}, \bibinfo {author} {\bibfnamefont
  {T.}~\bibnamefont {{Hoeksema}}}, \bibinfo {author} {\bibfnamefont
  {A.}~\bibnamefont {{Jim{\'e}nez}}}, \bibinfo {author} {\bibfnamefont {M.~F.}\
  \bibnamefont {{G{\'o}mez}}}, \bibinfo {author} {\bibfnamefont {J.~M.}\
  \bibnamefont {{Herreros}}}, \bibinfo {author} {\bibfnamefont {T.~R.}\
  \bibnamefont {{Cort{\'e}s}}}, \bibinfo {author} {\bibfnamefont {A.~R.}\
  \bibnamefont {{Jones}}}, \bibinfo {author} {\bibfnamefont {J.~M.}\
  \bibnamefont {{Pap}}}, \ and\ \bibinfo {author} {\bibfnamefont {R.~C.}\
  \bibnamefont {{Willson}}},\ }\href {\doibase 10.1007/BF00733428} {\bibfield
  {journal} {\bibinfo  {journal} {Solar Physics}\ }\textbf {\bibinfo {volume}
  {162}},\ \bibinfo {pages} {101} (\bibinfo {year} {1995})}\BibitemShut
  {NoStop}%
\bibitem [{\citenamefont {{Fr{\"o}hlich}}\ \emph {et~al.}(1997)\citenamefont
  {{Fr{\"o}hlich}}, \citenamefont {{Crommelynck}}, \citenamefont {{Wehrli}},
  \citenamefont {{Anklin}}, \citenamefont {{Dewitte}}, \citenamefont
  {{Fichot}}, \citenamefont {{Finsterle}}, \citenamefont {{Jim{\'e}nez}},
  \citenamefont {{Chevalier}},\ and\ \citenamefont {{Roth}}}]{VIRGO2}%
  \BibitemOpen
  \bibfield  {author} {\bibinfo {author} {\bibfnamefont {C.}~\bibnamefont
  {{Fr{\"o}hlich}}}, \bibinfo {author} {\bibfnamefont {D.~A.}\ \bibnamefont
  {{Crommelynck}}}, \bibinfo {author} {\bibfnamefont {C.}~\bibnamefont
  {{Wehrli}}}, \bibinfo {author} {\bibfnamefont {M.}~\bibnamefont {{Anklin}}},
  \bibinfo {author} {\bibfnamefont {S.}~\bibnamefont {{Dewitte}}}, \bibinfo
  {author} {\bibfnamefont {A.}~\bibnamefont {{Fichot}}}, \bibinfo {author}
  {\bibfnamefont {W.}~\bibnamefont {{Finsterle}}}, \bibinfo {author}
  {\bibfnamefont {A.}~\bibnamefont {{Jim{\'e}nez}}}, \bibinfo {author}
  {\bibfnamefont {A.}~\bibnamefont {{Chevalier}}}, \ and\ \bibinfo {author}
  {\bibfnamefont {H.}~\bibnamefont {{Roth}}},\ }\href {\doibase
  10.1023/A:1004929108864} {\bibfield  {journal} {\bibinfo  {journal} {Solar
  Physics}\ }\textbf {\bibinfo {volume} {175}},\ \bibinfo {pages} {267}
  (\bibinfo {year} {1997})}\BibitemShut {NoStop}%
\bibitem [{\citenamefont {{VIRGO Science Team}}(1995)}]{VIRGO3}%
  \BibitemOpen
  \bibfield  {author} {\bibinfo {author} {\bibnamefont {{VIRGO Science
  Team}}},\ }\href {https://www.ias.u-psud.fr/virgo/html/Purpleweb/purple.html}
  {\enquote {\bibinfo {title} {Virgo purple book, version 2.2},}\ } (\bibinfo
  {year} {1995})\BibitemShut {NoStop}%
\bibitem [{\citenamefont {Stone}\ \emph {et~al.}(1998)\citenamefont {Stone},
  \citenamefont {Frandsen}, \citenamefont {Mewaldt}, \citenamefont {Christian},
  \citenamefont {Margolies}, \citenamefont {Ormes},\ and\ \citenamefont
  {Snow}}]{ACE1}%
  \BibitemOpen
  \bibfield  {author} {\bibinfo {author} {\bibfnamefont {E.~C.}\ \bibnamefont
  {Stone}}, \bibinfo {author} {\bibfnamefont {A.}~\bibnamefont {Frandsen}},
  \bibinfo {author} {\bibfnamefont {R.}~\bibnamefont {Mewaldt}}, \bibinfo
  {author} {\bibfnamefont {E.}~\bibnamefont {Christian}}, \bibinfo {author}
  {\bibfnamefont {D.}~\bibnamefont {Margolies}}, \bibinfo {author}
  {\bibfnamefont {J.}~\bibnamefont {Ormes}}, \ and\ \bibinfo {author}
  {\bibfnamefont {F.}~\bibnamefont {Snow}},\ }\href@noop {} {\bibfield
  {journal} {\bibinfo  {journal} {Space Science Reviews}\ }\textbf {\bibinfo
  {volume} {86}},\ \bibinfo {pages} {1} (\bibinfo {year} {1998})}\BibitemShut
  {NoStop}%
\bibitem [{\citenamefont {{ACE Science Team}}(2004)}]{ACEdata}%
  \BibitemOpen
  \bibfield  {author} {\bibinfo {author} {\bibnamefont {{ACE Science Team}}},\
  }\href {http://www.srl.caltech.edu/ACE/ASC/level2/} {\enquote {\bibinfo
  {title} {Ace level 2 (verified) data},}\ } (\bibinfo {year}
  {2004})\BibitemShut {NoStop}%
\bibitem [{\citenamefont {{VIRGO Science Team}}(2018)}]{VIRGOdata}%
  \BibitemOpen
  \bibfield  {author} {\bibinfo {author} {\bibnamefont {{VIRGO Science
  Team}}},\ }\href {https://sohowww.nascom.nasa.gov/data/archive.html}
  {\enquote {\bibinfo {title} {The soho data archive},}\ } (\bibinfo {year}
  {2018})\BibitemShut {NoStop}%
\bibitem [{\citenamefont {{Excelitas Technologies}}(2020)}]{coverGlass}%
  \BibitemOpen
  \bibfield  {author} {\bibinfo {author} {\bibnamefont {{Excelitas
  Technologies}}},\ }\href
  {https://www.excelitas.com/product/space-qualified-cover-glass} {\enquote
  {\bibinfo {title} {Space-qualified cover glass},}\ } (\bibinfo {year}
  {2020})\BibitemShut {NoStop}%
\bibitem [{\citenamefont {Bender}\ \emph {et~al.}(1998)\citenamefont {Bender},
  \citenamefont {Brillet}, \citenamefont {Ciufolini}, \citenamefont {Cruise},
  \citenamefont {Cutler}, \citenamefont {Danzmann}, \citenamefont {Fidecaro},
  \citenamefont {Folkner}, \citenamefont {Hough}, \citenamefont {McNamara}
  \emph {et~al.}}]{LPPA}%
  \BibitemOpen
  \bibfield  {author} {\bibinfo {author} {\bibfnamefont {P.}~\bibnamefont
  {Bender}}, \bibinfo {author} {\bibfnamefont {A.}~\bibnamefont {Brillet}},
  \bibinfo {author} {\bibfnamefont {I.}~\bibnamefont {Ciufolini}}, \bibinfo
  {author} {\bibfnamefont {A.}~\bibnamefont {Cruise}}, \bibinfo {author}
  {\bibfnamefont {C.}~\bibnamefont {Cutler}}, \bibinfo {author} {\bibfnamefont
  {K.}~\bibnamefont {Danzmann}}, \bibinfo {author} {\bibfnamefont
  {F.}~\bibnamefont {Fidecaro}}, \bibinfo {author} {\bibfnamefont
  {W.}~\bibnamefont {Folkner}}, \bibinfo {author} {\bibfnamefont
  {J.}~\bibnamefont {Hough}}, \bibinfo {author} {\bibfnamefont
  {P.}~\bibnamefont {McNamara}},  \emph {et~al.},\ }\href@noop {} {\bibfield
  {journal} {\bibinfo  {journal} {Max-Planck Institut f{\"u}r Quantenoptik}\ }
  (\bibinfo {year} {1998})}\BibitemShut {NoStop}%
\bibitem [{\citenamefont {Danzmann}\ \emph {et~al.}(2011)\citenamefont
  {Danzmann}, \citenamefont {Prince}, \citenamefont {Binetruy}, \citenamefont
  {Bender}, \citenamefont {Buchman}, \citenamefont {Centrella}, \citenamefont
  {Cerdonio}, \citenamefont {Cornish}, \citenamefont {Cruise}, \citenamefont
  {Cutler} \emph {et~al.}}]{YellowBook2}%
  \BibitemOpen
  \bibfield  {author} {\bibinfo {author} {\bibfnamefont {K.}~\bibnamefont
  {Danzmann}}, \bibinfo {author} {\bibfnamefont {T.~A.}\ \bibnamefont
  {Prince}}, \bibinfo {author} {\bibfnamefont {P.}~\bibnamefont {Binetruy}},
  \bibinfo {author} {\bibfnamefont {P.}~\bibnamefont {Bender}}, \bibinfo
  {author} {\bibfnamefont {S.}~\bibnamefont {Buchman}}, \bibinfo {author}
  {\bibfnamefont {J.}~\bibnamefont {Centrella}}, \bibinfo {author}
  {\bibfnamefont {M.}~\bibnamefont {Cerdonio}}, \bibinfo {author}
  {\bibfnamefont {N.}~\bibnamefont {Cornish}}, \bibinfo {author} {\bibfnamefont
  {M.}~\bibnamefont {Cruise}}, \bibinfo {author} {\bibfnamefont {C.~J.}\
  \bibnamefont {Cutler}},  \emph {et~al.},\ }\href@noop {} {\bibfield
  {journal} {\bibinfo  {journal} {Assessment Study Report ESA/SRE}\ }\textbf
  {\bibinfo {volume} {3}} (\bibinfo {year} {2011})}\BibitemShut {NoStop}%
\bibitem [{\citenamefont {{Wright}}(1992)}]{Force}%
  \BibitemOpen
  \bibfield  {author} {\bibinfo {author} {\bibfnamefont {J.~L.}\ \bibnamefont
  {{Wright}}},\ }\href@noop {} {\emph {\bibinfo {title} {{Space sailing.}}}}\
  (\bibinfo {year} {1992})\BibitemShut {NoStop}%
\bibitem [{\citenamefont {Pedregosa}\ \emph {et~al.}(2011)\citenamefont
  {Pedregosa}, \citenamefont {Varoquaux}, \citenamefont {Gramfort},
  \citenamefont {Michel}, \citenamefont {Thirion}, \citenamefont {Grisel},
  \citenamefont {Blondel}, \citenamefont {Prettenhofer}, \citenamefont {Weiss},
  \citenamefont {Dubourg}, \citenamefont {Vanderplas}, \citenamefont {Passos},
  \citenamefont {Cournapeau}, \citenamefont {Brucher}, \citenamefont {Perrot},\
  and\ \citenamefont {Duchesnay}}]{scikit}%
  \BibitemOpen
  \bibfield  {author} {\bibinfo {author} {\bibfnamefont {F.}~\bibnamefont
  {Pedregosa}}, \bibinfo {author} {\bibfnamefont {G.}~\bibnamefont
  {Varoquaux}}, \bibinfo {author} {\bibfnamefont {A.}~\bibnamefont {Gramfort}},
  \bibinfo {author} {\bibfnamefont {V.}~\bibnamefont {Michel}}, \bibinfo
  {author} {\bibfnamefont {B.}~\bibnamefont {Thirion}}, \bibinfo {author}
  {\bibfnamefont {O.}~\bibnamefont {Grisel}}, \bibinfo {author} {\bibfnamefont
  {M.}~\bibnamefont {Blondel}}, \bibinfo {author} {\bibfnamefont
  {P.}~\bibnamefont {Prettenhofer}}, \bibinfo {author} {\bibfnamefont
  {R.}~\bibnamefont {Weiss}}, \bibinfo {author} {\bibfnamefont
  {V.}~\bibnamefont {Dubourg}}, \bibinfo {author} {\bibfnamefont
  {J.}~\bibnamefont {Vanderplas}}, \bibinfo {author} {\bibfnamefont
  {A.}~\bibnamefont {Passos}}, \bibinfo {author} {\bibfnamefont
  {D.}~\bibnamefont {Cournapeau}}, \bibinfo {author} {\bibfnamefont
  {M.}~\bibnamefont {Brucher}}, \bibinfo {author} {\bibfnamefont
  {M.}~\bibnamefont {Perrot}}, \ and\ \bibinfo {author} {\bibfnamefont
  {E.}~\bibnamefont {Duchesnay}},\ }\href@noop {} {\bibfield  {journal}
  {\bibinfo  {journal} {Journal of Machine Learning Research}\ }\textbf
  {\bibinfo {volume} {12}},\ \bibinfo {pages} {2825} (\bibinfo {year}
  {2011})}\BibitemShut {NoStop}%
\bibitem [{\citenamefont {{Rasmussen}}\ and\ \citenamefont
  {{Williams}}(2006)}]{GPref}%
  \BibitemOpen
  \bibfield  {author} {\bibinfo {author} {\bibfnamefont {C.~E.}\ \bibnamefont
  {{Rasmussen}}}\ and\ \bibinfo {author} {\bibfnamefont {C.~K.~I.}\
  \bibnamefont {{Williams}}},\ }\href@noop {} {\emph {\bibinfo {title}
  {{Gaussian Processes for Machine Learning}}}}\ (\bibinfo {year}
  {2006})\BibitemShut {NoStop}%
\bibitem [{\citenamefont {{Henry Faulks et.~al}}(2000)}]{FTR}%
  \BibitemOpen
  \bibfield  {author} {\bibinfo {author} {\bibnamefont {{Henry Faulks
  et.~al}}},\ }\href@noop {} {\emph {\bibinfo {title} {{Final Technical Report
  of the (Phase A) Study of the Laser Interferometer Space Antenna}}}},\
  \bibinfo {type} {Tech. Rep.}\ \bibinfo {number} {{Report No.~LI-RP-DS-009}}\
  (\bibinfo  {institution} {{LISA Project}},\ \bibinfo {year} {{2000}})\
  \bibinfo {note} {{ESTEC Contract no.~13631/99/NL/MS,
  https://lisa.nasa.gov/archive2011/Documentation/FTR.pdf}}\BibitemShut
  {NoStop}%
\bibitem [{\citenamefont {{Gizon}}\ \emph {et~al.}(2009)\citenamefont
  {{Gizon}}, \citenamefont {{Schunker}}, \citenamefont {{Baldner}},
  \citenamefont {{Basu}}, \citenamefont {{Birch}}, \citenamefont {{Bogart}},
  \citenamefont {{Braun}}, \citenamefont {{Cameron}}, \citenamefont {{Duvall}},
  \citenamefont {{Hanasoge}}, \citenamefont {{Jackiewicz}}, \citenamefont
  {{Roth}}, \citenamefont {{Stahn}}, \citenamefont {{Thompson}},\ and\
  \citenamefont {{Zharkov}}}]{DIS}%
  \BibitemOpen
  \bibfield  {author} {\bibinfo {author} {\bibfnamefont {L.}~\bibnamefont
  {{Gizon}}}, \bibinfo {author} {\bibfnamefont {H.}~\bibnamefont {{Schunker}}},
  \bibinfo {author} {\bibfnamefont {C.~S.}\ \bibnamefont {{Baldner}}}, \bibinfo
  {author} {\bibfnamefont {S.}~\bibnamefont {{Basu}}}, \bibinfo {author}
  {\bibfnamefont {A.~C.}\ \bibnamefont {{Birch}}}, \bibinfo {author}
  {\bibfnamefont {R.~S.}\ \bibnamefont {{Bogart}}}, \bibinfo {author}
  {\bibfnamefont {D.~C.}\ \bibnamefont {{Braun}}}, \bibinfo {author}
  {\bibfnamefont {R.}~\bibnamefont {{Cameron}}}, \bibinfo {author}
  {\bibfnamefont {T.~L.}\ \bibnamefont {{Duvall}}}, \bibinfo {author}
  {\bibfnamefont {S.~M.}\ \bibnamefont {{Hanasoge}}}, \bibinfo {author}
  {\bibfnamefont {J.}~\bibnamefont {{Jackiewicz}}}, \bibinfo {author}
  {\bibfnamefont {M.}~\bibnamefont {{Roth}}}, \bibinfo {author} {\bibfnamefont
  {T.}~\bibnamefont {{Stahn}}}, \bibinfo {author} {\bibfnamefont {M.~J.}\
  \bibnamefont {{Thompson}}}, \ and\ \bibinfo {author} {\bibfnamefont
  {S.}~\bibnamefont {{Zharkov}}},\ }\href {\doibase 10.1007/s11214-008-9466-5}
  {\bibfield  {journal} {\bibinfo  {journal} {Space Science Reviews}\ }\textbf
  {\bibinfo {volume} {144}},\ \bibinfo {pages} {249} (\bibinfo {year}
  {2009})},\ \Eprint {http://arxiv.org/abs/1002.2369} {arXiv:1002.2369
  [astro-ph.SR]} \BibitemShut {NoStop}%
\bibitem [{\citenamefont {Shaul}\ \emph {et~al.}(2006)\citenamefont {Shaul},
  \citenamefont {Aplin}, \citenamefont {Araujo}, \citenamefont {Bingham},
  \citenamefont {Blake}, \citenamefont {Branduardi-Raymont}, \citenamefont
  {Buchman}, \citenamefont {Fazakerley}, \citenamefont {Finn}, \citenamefont
  {Fletcher} \emph {et~al.}}]{shaul2006solar}%
  \BibitemOpen
  \bibfield  {author} {\bibinfo {author} {\bibfnamefont {D.}~\bibnamefont
  {Shaul}}, \bibinfo {author} {\bibfnamefont {K.}~\bibnamefont {Aplin}},
  \bibinfo {author} {\bibfnamefont {H.}~\bibnamefont {Araujo}}, \bibinfo
  {author} {\bibfnamefont {R.}~\bibnamefont {Bingham}}, \bibinfo {author}
  {\bibfnamefont {J.}~\bibnamefont {Blake}}, \bibinfo {author} {\bibfnamefont
  {G.}~\bibnamefont {Branduardi-Raymont}}, \bibinfo {author} {\bibfnamefont
  {S.}~\bibnamefont {Buchman}}, \bibinfo {author} {\bibfnamefont
  {A.}~\bibnamefont {Fazakerley}}, \bibinfo {author} {\bibfnamefont {L.~S.}\
  \bibnamefont {Finn}}, \bibinfo {author} {\bibfnamefont {L.}~\bibnamefont
  {Fletcher}},  \emph {et~al.},\ }in\ \href@noop {} {\emph {\bibinfo
  {booktitle} {AIP Conference Proceedings}}},\ Vol.\ \bibinfo {volume} {873}\
  (\bibinfo {organization} {American Institute of Physics},\ \bibinfo {year}
  {2006})\ pp.\ \bibinfo {pages} {172--178}\BibitemShut {NoStop}%
\end{thebibliography}%

\end{document}